\documentclass[twocolumn,prl,aps,showpacs,superscriptaddress,floatfix,showkeys,nofootinbib,10pt]{revtex4-1}
\usepackage{amsthm}
\usepackage{amsmath,latexsym,amssymb,verbatim,enumerate,graphicx}
\usepackage{color}
\usepackage{amsfonts,amssymb,amsmath}
\usepackage[]{graphics,graphicx,epsfig}
\usepackage{amsthm}
\usepackage{mathtools}

\usepackage[colorlinks=true,citecolor=blue,urlcolor=blue]{hyperref}
\usepackage{multirow}

\theoremstyle{remark}
\newtheorem{definition}{Definition} 
\newtheorem{prop}[definition]{Proposition}
\newtheorem{obs}[definition]{Observation}

\newtheorem{lemma}{Lemma}

\newcommand{\be}{\begin{equation}}
\newcommand{\ee}{\end{equation}}

\newcommand{\tr}{\text{tr}}

\DeclareMathOperator{\Tr}{Tr}

\newcommand{\ba}{\begin{aligned}}
\newcommand{\ea}{\end{aligned}}

\def\ben{\begin{eqnarray}}
\def\een{\end{eqnarray}}

\def\bei{\begin{itemize}}
\def\eei{\end{itemize}}

\begin{document}

\title{Steering is an essential feature of non-locality in quantum theory}

\author{Ravishankar Ramanathan}
\email{ravishankar.r.10@gmail.com}
\affiliation{Laboratoire d'Information Quantique, Universit\'{e} Libre de Bruxelles, Belgium}

\author{Dardo Goyeneche}
\affiliation{Institute of Theoretical Physics and Astrophysics, National Quantum Information Centre, Faculty of Mathematics, Physics and Informatics, University of Gda\'nsk, Wita Stwosza 57, 80-308 Gda\'nsk, Poland}
\affiliation{Institute of Physics, Jagiellonian University, Krak\'ow, Poland}
\affiliation{Faculty of Applied Physics and Mathematics, Gda\'{n}sk University of Technology, 80-233 Gda\'{n}sk, Poland}

\author{Sadiq Muhammad}
\affiliation{Department of Physics, Stockholm University, S-10691, Stockholm, Sweden}

\author{Piotr Mironowicz}
\affiliation{Institute of Theoretical Physics and Astrophysics, National Quantum Information Centre, Faculty of Mathematics, Physics and Informatics, University of Gda\'nsk, Wita Stwosza 57, 80-308 Gda\'nsk, Poland}
\affiliation{Department of Algorithms and System Modelling, Faculty of Electronics, Telecommunications and Informatics,
Gda\'{n}sk University of Technology, 80-233 Gda\'{n}sk, Poland}

\author{Marcus Gr\"unfeld}
\affiliation{Department of Physics, Stockholm University, S-10691, Stockholm, Sweden}

\author{Mohamed Bourennane}
\affiliation{Department of Physics, Stockholm University, S-10691, Stockholm, Sweden}

\author{Pawe{\l} Horodecki}
\affiliation{Institute of Theoretical Physics and Astrophysics, National Quantum Information Centre, Faculty of Mathematics, Physics and Informatics, University of Gda\'nsk, Wita Stwosza 57, 80-308 Gda\'nsk, Poland}
\affiliation{Faculty of Applied Physics and Mathematics, Gda\'{n}sk University of Technology, 80-233 Gda\'{n}sk, Poland}

\date{\today}

\begin{abstract}
\textbf{Abstract.}
A physical theory is called non-local when observers can produce instantaneous effects over distant systems. Non-local theories rely on two fundamental effects: local uncertainty relations and steering of physical states at a distance. In quantum mechanics, the former one dominates the other in a well-known class of non-local games known as XOR games. In particular, optimal quantum strategies for XOR games are completely determined by the uncertainty principle alone. This breakthrough result has yielded the fundamental open question whether optimal quantum strategies are always restricted by local uncertainty principles, with entanglement based steering playing no role. In this work, we provide a negative answer to the question, showing that both steering and uncertainty relations play a fundamental role in determining optimal quantum strategies for non-local games. Our theoretical findings are supported by an experimental implementation with entangled photons.
\end{abstract}

\maketitle

\textbf{Introduction.}
The uncertainty principle is a fundamental feature of quantum theory, which postulates the existence of incompatible observables, the results of whose measurements on identically prepared systems cannot be predicted simultaneously with certainty. Recently, the traditional formulation of uncertainty relations in terms of standard deviations and commutators has been eschewed in favor of the entropic uncertainty relations \cite{SW} and the even more fundamental fine-grained uncertainty relations \cite{OW}. These fine-grained uncertainty relations are formulated in terms of the basic entities of the theory, namely the probabilities of particular sets of outcomes for given sets of measurements, and are thus able to capture the uncertainty of these measurements in a more general manner than the entropic measures or the statistical standard deviations. Moreover, the uncertainty bounds are expressed in a manner independent of the specific underlying quantum state, an advantage over the traditional formulation in terms of average values of commutators on fixed states. Another fundamental feature of quantum theory is steering, identified by Schr\"{o}dinger in \cite{Schrodinger}. This property determines, for two systems in a shared (entangled) state, which states can be prepared on one system by a measurement on the other. Quantum steering can be used as a resource to generate ensambles of quantum systems incompatible with a local hidden variable (LHV) model \cite{Wiseman}. For two-qubit states, all states that Alice can steer are restricted to an ellipsoid within the Bloch sphere of Bob \cite{Jevtic}.

The results of measurements on distant quantum systems can be correlated in a way that defies classical local realistic description. This non-locality of quantum theory is evidenced in the violation of Bell inequalities by spatially separated quantum systems. Quantum correlations are restricted to some extent by the no-signaling principle, i.e., the measurement results cannot allow for signaling between the distant locations. Nevertheless, there exist non-local correlations allowed by the no-signaling principle that cannot be realized in quantum theory \cite{PR, our}.

The fundamental question why quantum correlations are non-local yet not as strong as allowed by the no-signaling principle is an intriguing one that has stimulated the formulation of many striking new information-theoretic principles. So far none of the known principles has been able to capture the set of quantum correlations in its entirety \cite{AQ}, thus a comprehensive answer to this question is still lacking. 
The test-beds for these principles are a special class of Bell inequalities based on so-called quantum non-local games which extract purely probabilistic aspects of the non-locality test, independent of the physical realization. Consideration of non-local games lead to a  significant  breakthrough  in \cite{OW}  where  two  fundamental  concepts  of  quantum
theory, the strength of non-local correlations and the uncertainty principle, were shown to be inextricably quantitatively linked with each other.

Moreover it was shown that in a large class of non-local games for which optimal quantum strategies were explicitly known (the class of XOR games for which an explicit characterization of the optimal quantum strategy was provided by Tsirelson \cite{Tsirelson}) these are not only just linked, but one of them -  uncertainty -  fully determines the non-locality of quantum theory with steering playing no role. An important question left open in \cite{OW} was whether such a phenomenon holds in general. If it did, this would constitute a defining property of quantum
mechanics: that something fully local (the uncertainty principle for a single party's measurements)
governs something non-local (the Bell violation on a shared system). 

The intriguing result of \cite{OW} is that while the degree of non-locality in any theory is generally determined by a combination of two factors - the strength of the uncertainty principle and the degree of steering allowed in the theory, in quantum theory the degree of non-locality for the well-known class of two-player XOR games  is purely determined by the strength of the uncertainty principle alone. More precisely, \cite{OW} shows that in a two-party Bell scenario, the strength of non-locality in any theory is determined by the uncertainty relations for Bob's measurements acting on the states that Alice can steer to. On the other hand in quantum theory, for all XOR games (aka bipartite correlation Bell inequalities) \cite{Tsirelson},
the states which Alice can steer to are identical to the most certain states, so that only the uncertainty relations of Bob's local measurements determine the outcome. 

In this paper, we show that the one-to-one correspondence between the uncertainty principle and the degree of non-locality in quantum theory (referred hereafter as the Uncertainty Principle - Quantum Game Value correspondence, or UP-QGV correspondence) observed for XOR games in \cite{OW} does not hold in general, by presenting an explicit counter-example of a non-local game violating the correspondence. We provide an intuitive explanation in terms of the Schrodinger-Hughston-Jozsa-Wootters theorem \cite{HJW} for when the UP-QGV correspondence breaks down. To show that the game does not have other optimal strategies that could obey the correspondence and to facilitate experimental testing of our result, we prove a self-testing property of the game, namely that there is a unique state and measurements (up to local unitaries and attaching irrelevant ancillae) that achieves the optimal quantum value. Furthermore, the game is not an isolated example, we extend it to show that every two-party non-maximally entangled state $|\psi \rangle$ is the optimal state for a game $G_{\psi}$ for which the correspondence does not hold. The tradeoff existing between steering and uncertainty is conclusively shown by means of an experimental implementation, in which the steered states manifestly are seen to be distant from the maximally certain state even after the experimental errors are taken into account.

\textbf{Results}

\textbf{Uncertainty Principle - Quantum Game Value correspondence.}
Let us first recall the precise correspondence between the fine-grained uncertainty relations and the strength of non-locality established in \cite{OW}. Consider a two-player non-local game $G$, in which Alice and Bob receive questions $x, y$ from respective input sets $\text{X}, \text{Y}$ according to some input distribution $\pi_\text{X,Y}(x,y)$. They return answers $a, b$ from some output sets $\text{A}, \text{B}$, respectively.
The winning constraint 
is specified by a predicate $V(a,b|x,y) \in \{0,1\}$. The success probability in the game $\omega_\text{s}(G)$ 
is thus written as
\begin{eqnarray}
\label{eq:game-val}
\omega_\text{s}(G) =&& \max_{P_\text{A,B$|$X,Y} \in \mathcal{S}}	\sum_{\substack{x \in \text{X} \\ y \in \text{Y}}} \pi_\text{X,Y}(x,y) \nonumber \\
 &&\qquad \qquad\sum_{\substack{a \in \text{A} \\ b \in \text{B}}} V(a,b|x,y) P_\text{A,B$|$X,Y}(a,b|x,y), 
\end{eqnarray}
where $\mathcal{S}$ refers to a set of conditional probability distributions (boxes) $P_\text{A,B$|$X,Y}$. 
One considers boxes taken from sets $\text{C}, \text{Q}, \text{NS}$ corresponding to the set of classical, boxes and general no-signaling boxes, with corresponding values $\omega_\text{c}(G), \omega_\text{q}(G)$ and $\omega_\text{ns}(G)$ respectively. One may also restrict attention to the free games for which the input distributions are independent, i.e., $\pi_\text{X,Y}(x,y) = \pi_\text{X}(x) \pi_\text{Y}(y)$.

We will in particular be interested in $\omega_\text{q}(G)$, i.e., the value obtained from those boxes for which there exists a state $\rho$ on a Hilbert space $\mathcal{H}_d$ and sets of measurement operators (POVMs) $\{M^{x}_{a} \}, \{M^{y}_{b}\}$ such that
	$P_\text{A,B$|$X,Y}(a,b|x,y) = \Tr \left(\rho M^{x}_{a} \otimes M^{y}_{b} \right)$.
The idea in \cite{OW} is to rewrite the game expression in Eq.(\ref{eq:game-val}) as
\be
	\label{eq:game-val-2}
	\sum_{\substack{x,  a}} \pi_\text{X}(x) P_\text{A$|$X}(a|x) \sum_{y, b} \pi_\text{Y$|$X}(y|x) V(a,b|x,y) P_\text{B$|$Y,X,A}(b|y,x,a). 
\ee
Let $P_\text{B$|$Y,X,A}(b|y, x,a)_{\hat{\sigma}^{B}_{a|x}}$ be Bob's marginal probability distribution when his state is steered by Alice to $\hat{\sigma}^{B}_{a|x}$. Now, observe that for each $(x,a)$, the expression 
\be 
	\label{eq:game-val-uncert}
	\sum_{y, b} \pi_{Y}(y) V(a,b|x,y) P_\text{B$|$Y,X,A}(b|y, x,a)_{\hat{\sigma}^{B}_{a|x}} \leq \xi_{B}^{(x,a)}, 
\ee
constitutes a fine-grained uncertainty relation on Bob's system with $\xi_{B}^{(x,a)}$ denoting the maximum over all possible states $\hat{\sigma}^{B}_{a|x}$ of Bob's system. When the optimal value $\xi_B^{(x,a)}$ equals unity, we refer to the corresponding uncertainty relation as trivial,
i.e., while the probabilities are bounded below unity for some states, there exist states for which the outcomes (for each of Bob's inputs $y$) can be fixed with certainty. On the other hand, when $\xi_{B}^{(x,a)} < 1$, we infer that one cannot obtain a measurement outcome with certainty for all measurements simultaneously.

\begin{figure}[ht]
	\includegraphics[scale=0.1]{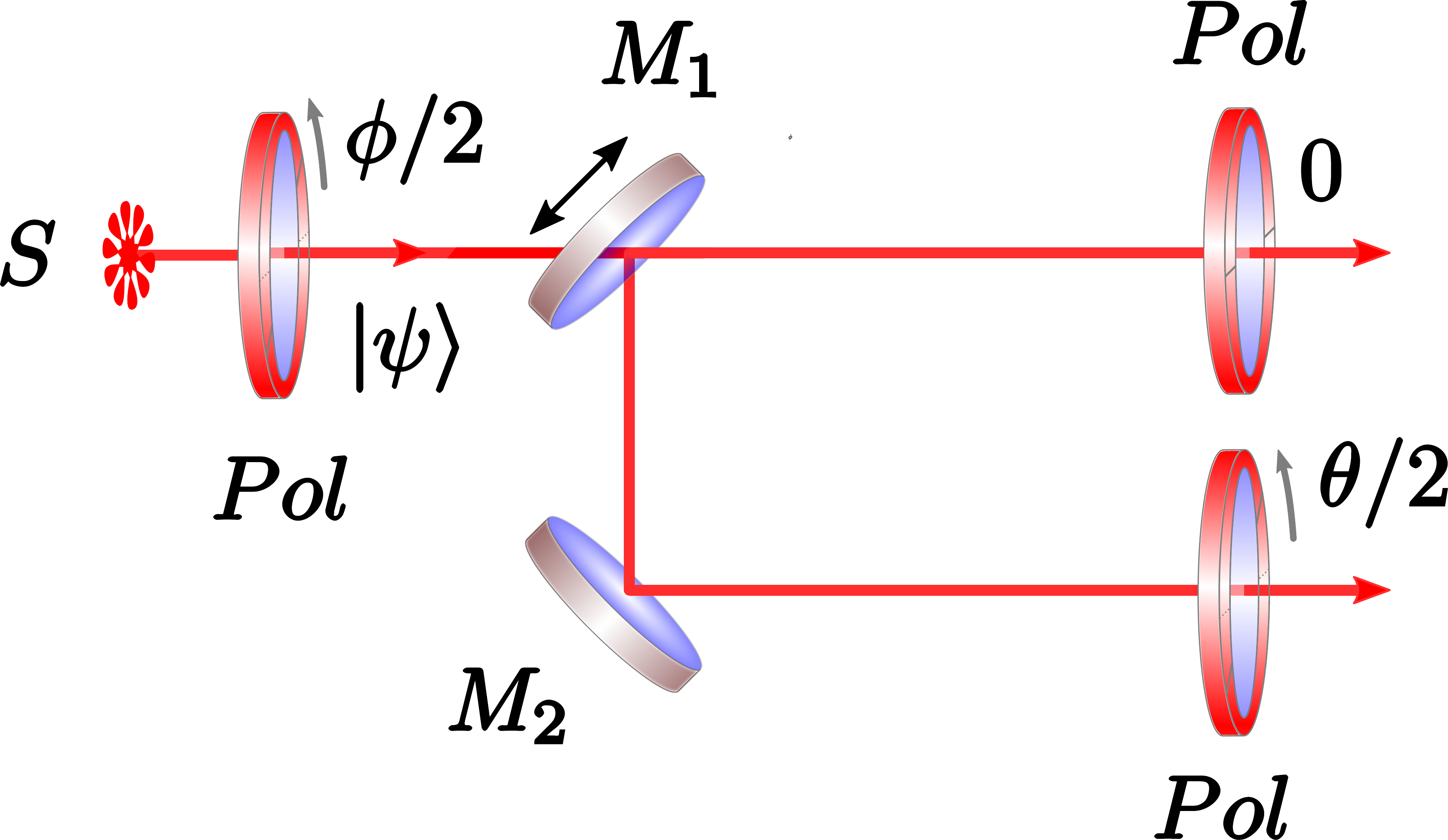}
	\caption[The uncertainty principle illustration]{
	The uncertainty principle illustrated by randomly oriented 	polarizers. Input state $|\psi\rangle$ is prepared via a polarizer (Pol) oriented at $\phi/2$, (which corresponds to orientation $\phi$ on the Bloch sphere). A reflecting mirror $M_1$ is randomly inserted with probability $0<p<1$ in the path of the photons. A polarizer at $0$ measures observable $Q(0)$, and another one rotated by $\frac{\theta}{2}$ ($0 < \theta < \pi$) measures $Q(\theta)$, such that probability that a photon is transmitted, is   $P(transmission) = (1-p) Q(0)_{|\psi\rangle}+ p Q(\theta)_{|\psi\rangle}$ and it is upper bounded by $\xi(\theta,p)$.
	}
	\label{uncertainty1}
\end{figure}

\begin{figure}[ht]
	\includegraphics[width=7cm]{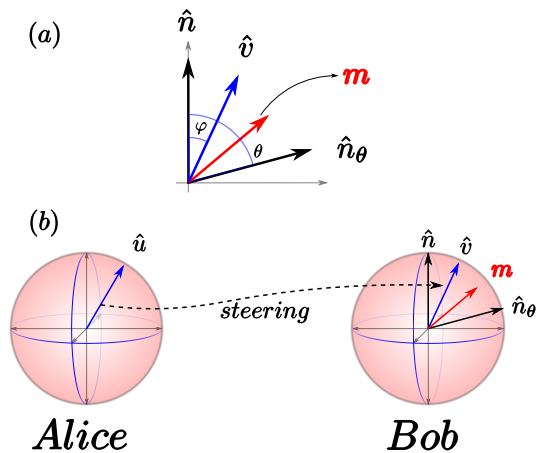}
	\caption[The Bloch sphere representation of the measurement]{(a) The Bloch sphere representation of the measurement situation. The state $| \psi \rangle$ of the polarized photon is represented by $\hat{v}$, while the projectors $Q(0)$ and $Q(\theta)$ correspond to unit vectors $\hat{n}$ and $\hat{n}_{\theta}$ respectively, and $m$ is give by $m =(1-p) \hat{n} + p \hspace{0.5mm} \hat{n}_{\theta} $. The bound on the probability of transmission $\xi(\theta,p)$
		is obtained from the vector $m$,  $\xi(\theta,p)=\frac {1+|m|}{2}$.
		The uncertainty relation defined by the probability of transmission ( $P(\mbox{transmission}) \leq \xi(\theta,p) < 1$ ) is saturated by the $| \psi \rangle$ with Bloch vector $\hat{v}$ parallel to $m$.
		(b) The situation when Alice tries to steer to the least uncertain state. It is achieved  only when  $\hat{v} \Vert{m}$.}
	\label{uncertainty2}
\end{figure}

An example situation of the uncertainty relation is shown in Fig. (\ref{uncertainty1}) and steering to the maximally certain states is exemplified in Fig. (\ref{uncertainty2}). 

Let $\{\tilde{\sigma}^{B}_{a|x} \}$ denote the set of states of Bob's system that achieve the maximum value $\xi_{B}^{(x,a)}$ of the uncertainty expressions for each $(x,a)$ for given optimal measurement operators $\{ M^{y}_{b} \}$. The question then arises whether Alice is able to steer Bob's system to these maximally certain states and thus achieve the bound set by the uncertainty principle for the game $G$. We are thus lead to consider the effect of steering. For any bipartite no-signaling box shared by Alice and Bob, any measurement on Alice's system creates a set of single-party boxes on Bob's side $\{P_\text{B$|$Y}(b|y)_{x,a} \}=\{ P_\text{B$|$Y,X,A}(b|y,x,a) \}$. We say that with this particular input-output pair $(x,a)$, Alice has steered the state of Bob's system to the set of boxes $\{P_\text{B$|$Y}(b|y)_{x,a}\}$ with probability $P_\text{A$|$X}(a|x)$. 

We see therefore in Eq.(\ref{eq:game-val-2}) the separation of the game expression into two components, one where Bob's (optimal) measurements define a set of uncertainty relations one for each $(x,a)$ and a second component wherein Alice tries to steer Bob's system to the maximally certain states for these relations. The strength of non-locality in any theory is thus seen as a trade-off between the strength of the uncertainty relations and the amount of steering allowed in the theory. 


In \cite{OW}, it was shown that for the well-known class of two-player XOR games for which the optimal measurements are known, the strength of non-locality is purely determined by the uncertainty relation with steering not constraining the value in any way. In other words, the optimal measurements and the state share the property that in all these known instances, Alice is able to steer Bob's system to the most certain states corresponding to the set of uncertainty relations of his system for each input-output pair $(x,a)$. 

Note that the restriction to non-local games rather than all Bell inequalities is crucial for the correspondence to be meaningful. Indeed, for general Bell inequalities, where one is allowed to scale the Bell expression with arbitrary multiples of the normalization and no-signaling equalities, it is possible to show that the correspondence can always be made to hold up to arbitrary high accuracy. This general observation inspired by recent results in \cite{Scarani} is explained in detail in the Supplementary Note 3. 

Two-player XOR games are non-local games with an arbitrary number of inputs and binary outputs, where the winning constraint of the game only depends on the \textsc{xor} of the parties' outputs. Building on a breakthrough theorem by Tsirelson \cite{Tsirelson}, it was shown in \cite{Wehner06, Cleve} that the quantum value of two-party \textsc{xor} games can be calculated precisely by means of a semi-definite program, and the Tsirelson theorem allows to recover the optimal state and measurement operators for any such game. In effect, apart from the pseudo-telepathy games \cite{Brassard} and a few other isolated instances, these are the games where the optimal measurements are known and for which the relation between the uncertainty principle and non-locality was established in \cite{OW}. The difficulty in establishing the relationship for general non-local games is due to the fact that the problem of finding the quantum strategy of arbitrary non-local games is hard \cite{Kempe}; one usually uses a hierarchy of semi-definite programs \cite{NPA, NPA-dual} which converge to the true quantum value.

Note that it is natural to ask about the relation of the steering-type representation of the formula (\ref{eq:game-val-2}) to the well-known 
Schrodinger-Hughston-Jozsa-Wootters theorem which defines all the ensembles Alice may steer to. It is tempting to expect that 
the result \cite{OW} is due to application of that theorem. 
This is however {\it not} the case because of the crucial fact 
it is not guaranteed that the maximally certain states
together with the optimal local probabilities $P_\text{A$|$X}(a|x)$ obey the no-signaling condition of the SHJW theorem (see Supplementary Note 2 for more discussion), viz.
\be 
\label{eq:steering-ns}
\sum_{a} P_\text{A$|$X}(a|x) \hat{\sigma}^\text{B}_{a|x} = \sum_{a} P_\text{A$|$X}(a|x') \hat{\sigma}^\text{B}_{a|x'} = \hat{\sigma}_{B} = \tr_{A} \hat{\sigma}_{AB}.
\ee 
   
\textbf{Counter-examples to the correspondence.}
Let us now exhibit an example of a non-local game for which the
UP-QGV correspondence does not hold, i.e., one
where the optimal quantum state and measurements are such that Alice is unable to steer Bob's system to the maximally certain state for each $(x,a)$. Before we proceed to the counter-example, let us mention that it is possible that the optimal quantum value of a non-local game can be achieved with different sets of states and measurement operators (even going beyond a trivial unitary equivalence), therefore one must check whether the relation could hold for at least one optimal quantum strategy. Thus, in order to give a counter-example to the UP-QGV correspondence, it is necessary to prove that the relationship does not hold for all optimal quantum strategies for the game. We achieve this requirement by proving a self-testing property of the counter-example, i.e., that up to unitary equivalences there is a unique state and sets of measurements that achieves the optimal value of the game. 


We consider the Bell scenario $\textit{B}(2,2,2)$ of two parties, each performing one of two measurements and obtaining one of two outputs.
The Bell inequality corresponding to the game denoted $G^{(7)}$ is explicitly given by
\be
\label{eq:Bell-exp-7}
	\ba
\frac{1}{4} [P & (0,0|0,0) + P(1,1|0,0) + P(0,1|0,1) + P(1,0|0,1)\\
		& + P(0,1|1,0) + P(1,0|1,0) + P(0,1|1,1) ] \leq \frac{3}{4},
	\ea
\ee
where we have assumed that each party chooses their inputs uniformly, i.e., $\pi_X(x) = \pi_Y(y) = \frac{1}{2}$ for $x, y \in \{0,1\}$ so that $\pi_{X,Y}(x,y) = \frac{1}{4}$ and the classical bound is $\omega_c(G^{(7)}) =\frac{3}{4}$. The optimal strategy for the game $G^{(7)}$ violates the UP-QGV correspondence (the proof of the following Proposition~\ref{prop-game-1} is given in the Supplementary Note 1). 

\begin{prop}
\label{prop-game-1}
The optimal quantum strategy for the game $G^{(7)}$ (achieving $\omega_\text{q}(G^{(7)}) \approx 0.782$) violates the uncertainty principle - quantum game value correspondence, i.e., Alice is unable to steer Bob's system to the maximally certain states and vice versa.
\end{prop}

The uncertainty relations for each input-output pairs $(x,a)$ of Alice for the game $G^{(7)}$ are given as
\begin{eqnarray}
\label{eq:uncertainty-game2-text}
&&(x=0,a=0) \rightarrow \nonumber \\
	&& \quad	P_\text{B$|$Y}(b=0|y=0) + P_\text{B$|$Y}(b=1|y=1) \leq 2 \xi_B^{(0,0)} \nonumber \\
&&(x=0,a=1) \rightarrow \nonumber \\
	&&	\quad P_\text{B$|$Y}(b=1|y=0) + P_\text{B$|$Y}(b=0|y=1) \leq 2 \xi_B^{(0,1)} \nonumber \\
&&(x=1,a=0) \rightarrow \nonumber \\
	&&	\quad P_\text{B$|$Y}(b=1|y=0) + P_\text{B$|$Y}(b=1|y=1) \leq 2 \xi_B^{(1,0)} \nonumber \\
&&(x=1,a=1) \rightarrow P_\text{B$|$Y}(b=0|y=0) \leq 1,
\end{eqnarray}
where
the uncertainty bounds are $\xi_B^{(0,0)} = \xi_{B}^{(0,1)} \approx 0.882$, 
and $\xi_B^{(1,0)} \approx 0.823$. The optimal state and measurements achieving $\omega_\text{q}(G^{(7)}) \approx 0.782$ are given in the Supplementary Note 4, where it is shown explicitly that while for $(x=1,a=0)$ Alice steers Bob's system to the maximally certain state, for $(x=0,a=0)$ and $(x=0,a=1)$ Alice is unable to steer Bob's system to the maximally certain states of the corresponding (non-trivial) uncertainty relations. Further, the trivial uncertainty relation for $(x=1,a=1)$ also fails to be saturated. The value $\omega_\text{q}(G^{(7)})$ achievable in quantum theory is thus strictly lower than what is allowed by the uncertainty principle, and therefore the game $G^{(7)}$ violates the UP-QGV correspondence. 

Let us now see why the UP-QGV correspondence breaks down for the particular game $G^{(7)}$, and establish conditions for the correspondence to hold. To do so, we examine the assemblage $\{P_\text{A$|$X}(a|x), \mathbf{\tilde{\sigma}}_{a|x}\}$ of maximally certain states. 
For the game $G^{(7)}$ it can be readily verified that the corresponding assemblage of maximally certain states does not obey the no-signaling relation Eq.(\ref{eq:steering-ns}), so the SHJW theorem does not guarantee the existence of a shared entangled state and measurements on Alice's side that would prepare the corresponding maximally certain states on Bob's system. Formally, we may make the observation (which follows from well-known demands on steerability \cite{Wiseman}) that the UP-QGV correspondence holds when the probabilities $P_\text{A$|$X}(a|x)$ together with the maximally certain states $\hat{\sigma}^\text{B}_{a|x}$ obey the no-signaling constraint in Eq.(\ref{eq:steering-ns}).
\begin{obs}
	\label{obs1}
	The uncertainty principle determines the non-locality of quantum theory whenever the maximally certain states $\hat{\sigma}^\text{B}_{a|x}$ of one party's measurements together with the optimal local probabilities $\{P(a|x)\}$ of the other party, forms a no-signaling assemblage, i.e., when $\{P(a|x), \hat{\sigma}^\text{B}_{a|x}\}$ obeys Eq.(\ref{eq:steering-ns}).
\end{obs}
The game $G^{(7)}$ shows that this condition is not always obeyed by the maximally certain states. While it appears at present an intractable problem to characterize the set of all games where the UP-QGV correspondence breaks down, we can nevertheless show that the game $G^{(7)}$ is not singular in this respect. Indeed, every two-party non-maximally entangled state $| \psi \rangle$ (i.e. a state not of the form $\frac{1}{\sqrt{d}}\sum_{i=1}^{d} |i, i \rangle$ for some $d > 1$) is the optimal state for a game $G_{\psi}$ for which the correspondence does not hold. This is captured in the following proposition (whose proof is given in the Supplementary Note 4). 

\begin{prop}
	\label{prop-game-2}
	For any two-party entangled, but non-maximally entangled, state $| \psi \rangle \in \mathbb{C}^d \otimes \mathbb{C}^d$ for arbitrary Hilbert space dimension $d$, there exists a game $G_{\psi}$ for which the optimal quantum strategy is given by suitable measurements on $| \psi \rangle$, and such that the correspondence between the uncertainty principle and the quantum game value does not hold for $G_{\psi}$.  
\end{prop}

An interesting open question is whether the conditions in Observation \ref{obs1} are met for all unique games \cite{Kempe2} which are a natural generalization of XOR games to a larger output alphabet. Also interesting is to find whether the correspondence holds for all games where the optimal strategy involves a maximally entangled state, which would highlight that in the foundational program of seeking an information-theoretic principle behind the strength of quantum non-local correlations, one must go further than the correlations exhibited by the maximally entangled states alone.

\textbf{Experimental Implementation.}
In our experiment, the physical qubits are  single  photon polarization states and the computational basis corresponds to the horizontal (H) and vertical (V) polarization i.e. $|H\rangle \equiv |0\rangle$ and $|V\rangle \equiv |1\rangle$. To achieve the maximal violation of the Bell inequality given in \eqref{eq:Bell-exp-7}, we used the following polarization entangled two-photon state,
\begin{eqnarray}{\label{eqE_1}}
|\Psi\rangle &=& 0.2487 |HH\rangle +  0.4760 |HV\rangle  \\ \nonumber
&&+  0.8060 |VH\rangle -  0.2487 |VV\rangle.
\end{eqnarray}
This state is produced in two steps. First, we generate entangled photon pairs via spontaneous parametric down-conversion (SPDC) \cite{SPDC_cross}. Then at the second step, these entangled pairs are transformed to the required state \eqref{eqE_1} by local rotations \cite{State_prep}. The polarization measurement on Alice and Bob's sides are performed by analyzers consisting of wave-plates, polarizing beam splitters (PBS) and single photon detectors. An FPGA based timing system is used to collect data. The experimental setup is outlined in Fig.~\ref{fig1} and its detailed description can be found in the Supplementary Note 5.

\begin{figure}[tb]
	\begin{center}
		\centerline{\includegraphics[width=1\columnwidth]{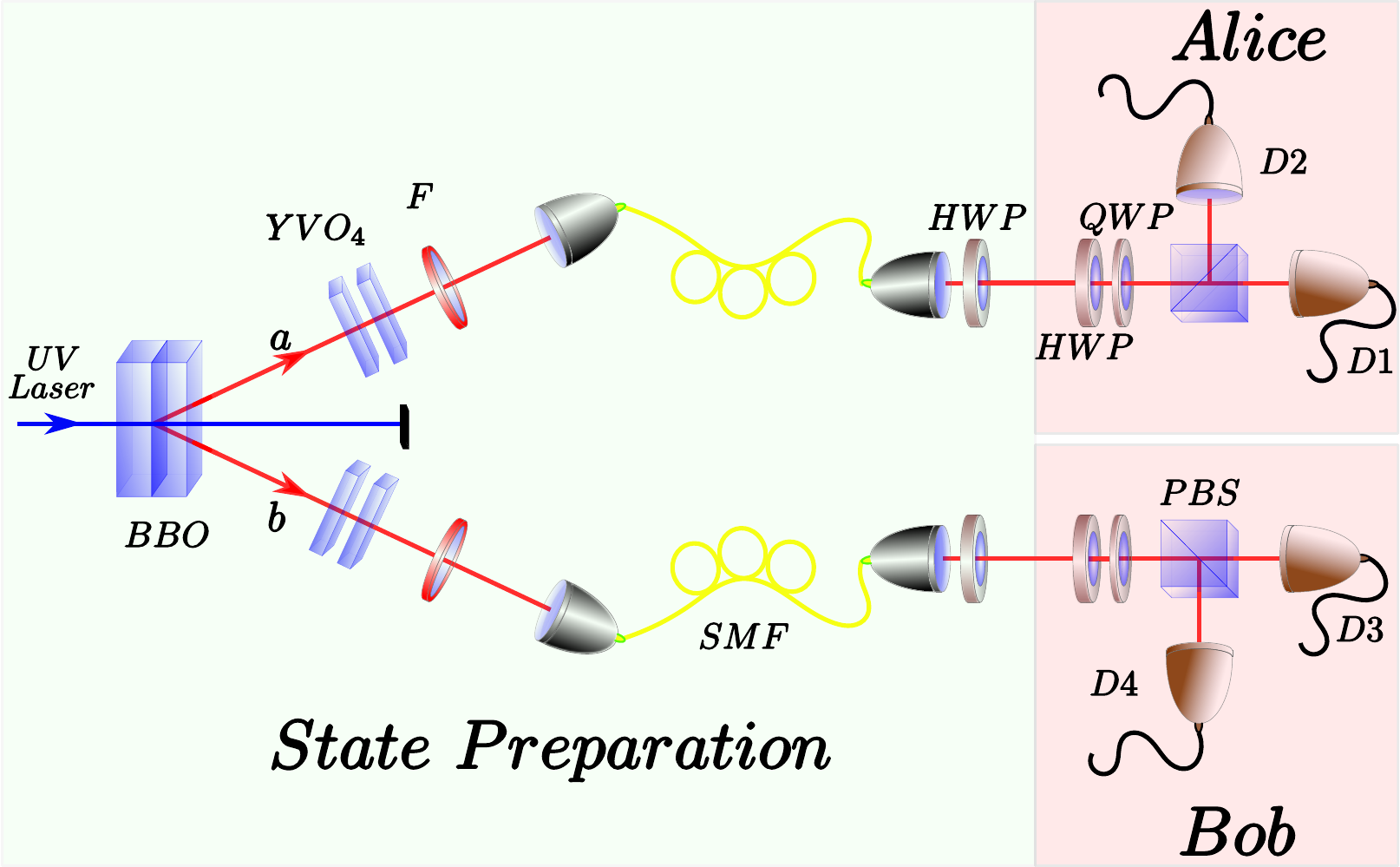}}
		\caption[Experimental setup.]{(Color online) Preparation and measurement stages for the state \eqref{eqE_1}. A UV pump laser at 390\,nm was focused onto two $\beta$-barium borate (BBO) crystals placed in cross-configuration to produce photon pairs emitted into two spatial modes $``a"$ and $``b"$ through type-I SPDC process. Any spatial, temporal or spectral distinguishability between the photons is removed via a pair of $YVO_4$ crystals, narrow-bandwidth filters (F) and coupling into single mode fibers (SMF). Then, the photons in each mode are rotated through a half wave-plate to get the desired state \eqref{eqE_1}. For measurement, Alice and Bob uses polarization analyzers consisting of a half wave-plate (HWP), a quarter wave-plates (QWP), a polarizing beam splitter (PBS) and $D_i$ $(i=\{1,2,3,4\})$ single photon avalanche photo-diodes. }
		\label{fig1}
	\end{center}
\end{figure}

\begin{figure}[!h]
	\centering 
	{\includegraphics[width=7cm]{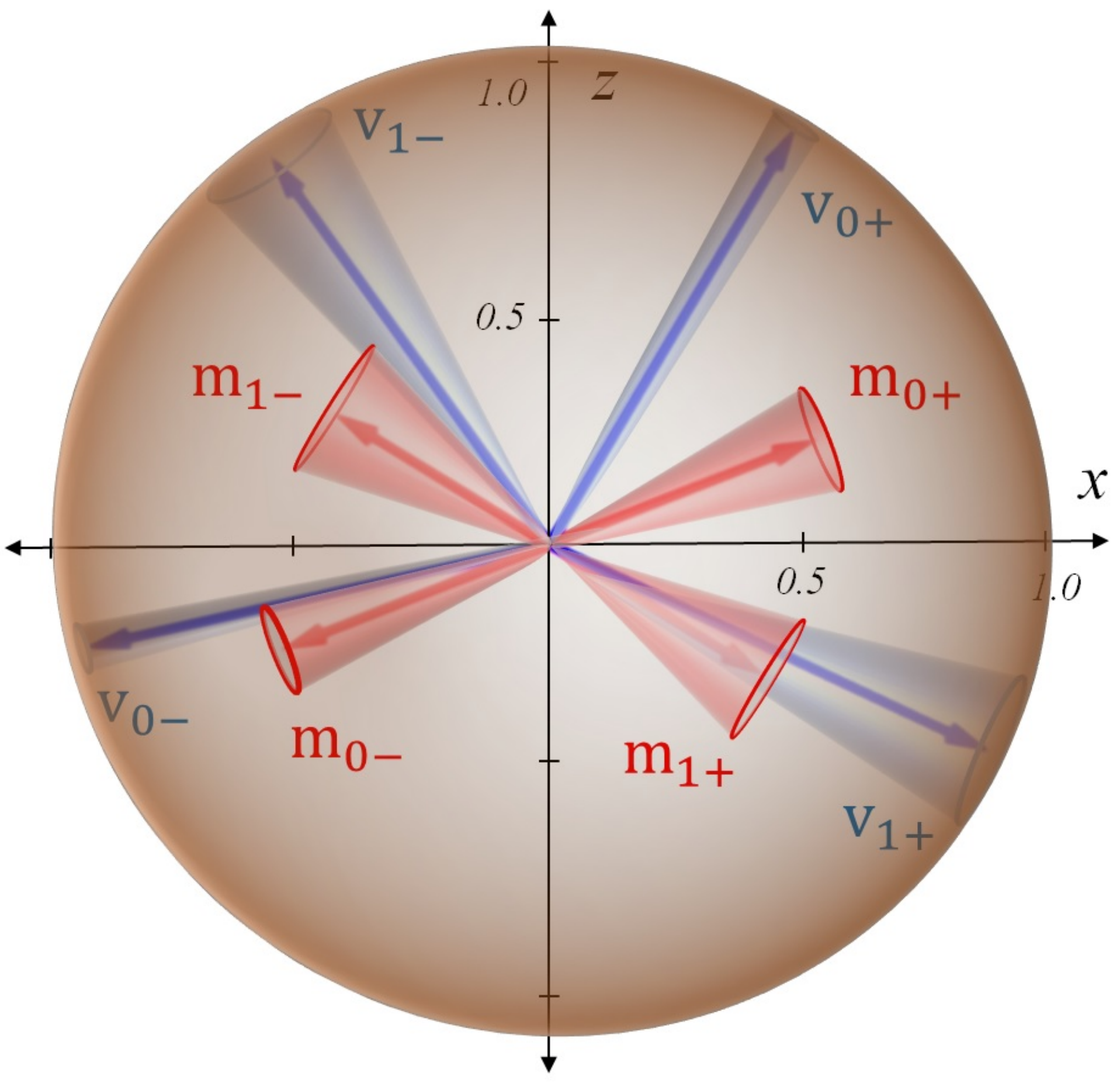}} 
	\caption[Experimental results]{(color online). Experimental results. Least uncertain states $\mathrm{v}$ (red) and states $\mathrm{m}$ that Alice is able to steer (blue). Cones show experimental errors originating from statistics (Poissonian) and systematic due to limited precision of the settings and non-ideal components. The experimental results illustrate that steering to the maximally certain state is not possible, as cones associated to $\mathrm{v}_{0+}$ and  $\mathrm{m}_{0+}$ do not intersect.}
	\label{Fig4}
\end{figure}

The Fidelity, $F=\langle\Psi|\rho_{exp}|\Psi\rangle$, of the experimentally prepared state $\rho_{exp}$ with respect to \eqref{eqE_1} was $0.9933 \pm 0.0009$. With this state and using the settings $|\phi_x^{\pm}\rangle = \cos\gamma_x |H\rangle \pm \sin\gamma_x |V\rangle$, where $\gamma_0=\pi/4$ and $\gamma_1=4.7948$, we obtained the experimental Bell inequality violation $\omega_\text{q}(G^{(7)})=0.7770\pm 0.0002$. Note that  the theoretical quantum and classical bounds are  $0.7822$ and $0.7500$ respectively. The fidelity of the four maximally certain states $v_{0+},v_{0-},v_{1+}$ and $v_{1-}$ are given by $F_{0+} = 0.9990 \pm 0.0003$, $F_{0-} = 0.9888 \pm 0.0008$, $F_{1+} = 0.9899 \pm 0.0009$ and $F_{1-} = 0.9957 \pm 0.0004$, respectively. Here, $v_{ij}$ is the least uncertain state associated to Alice measurement $i$ having outcome $j$. In Fig. \ref{Fig4} we represent the least uncertain states (blue) and the states $m_{ij}$ that Alice is able to steer (red) (see Supplementary Notes 3 and 5 for details related to theoretical and experimental results, respectively). Experimental errors determine eight cones in Bob's Bloch sphere, whose apertures are the largest possible, according to the experimentally obtained errors.

For error estimation, we have considered the error originated from the measurement side only, as the error on the preparation side will just shift the experimentally prepared state away from the desired state and therefore will be apparent from the reported state fidelity or the value of Bell violation. Further details are given in Supplementary Note 8.

We note that the experimental realization is not strictly required for the case of the paper. However, it is fundamental to note that the breakdown in the correspondence between the two major aspects of quantum theory is not a trivial one that would be washed out under inevitable experimental error, since the correspondence was only considered for the optimal quantum value. As such, it is of interest to find that even with current experimental technology, one can achieve sufficient experimental fidelities to make the case of the paper, apart from serving as one of the first experiments to self-test a non-maximally entangled state. Finally, we remark that the experiment was not performed in a loophole-free manner, as such it would be interesting to check the expectation that the same conclusions also hold in a loophole-free Bell test such as recently done in \cite{loophole-free1,loophole-free2,loophole-free3}.

\vspace{0.5cm}

\textbf{Discussion}

In this paper, we have shown that the intriguing correspondence between the uncertainty principle and the quantum game value, proven for the very important class of two-player XOR games in \cite{OW}, does not hold for general non-local games.
In order to prove this result we have put forth an intuitive argument to identify when the correspondence holds in terms of the SHJW theorem.

Many interesting questions remain open.
Firstly, note that the CHSH inequality is the only facet-defining inequality in the Bell scenario $B(2,2,2)$ and the non-local game we consider constitutes a lower-dimensional face of the classical polytope.
It is of interest to find whether the correspondence holds for non-local games that are tight Bell inequalities (facets of the classical polytope), or for games where the optimal strategy involves a maximally entangled state. Secondly, while the uncertainty relations always provide a bound on the quantum value, it is now an open question to characterize the class of games for which this bound is saturated and more interestingly those for which the gap is extremal.

 \vspace{0.5cm}
 
\textbf{Data availability statement.}
The data that support the findings of this study are available from the corresponding author upon reasonable request.

\textbf{Acknowledgments.}
R.R. thanks Nicolas Brunner and Antonio Ac\'{i}n for useful discussions. The paper is supported by ERC AdG grant QOLAPS, by Foundation for Polish Science TEAM project co-financed by the EU European Regional Development Fund, by the project ''Causality in quantum theory: foundations and applications" of the Fondation Wiener-Anspach
and the Interuniversity Attraction Poles 5 program of the Belgian Science Policy Office under the grant IAP P7-35 photonics@be, by the National Science Centre (NCN) grant 2014/14/E/ST2/00020, by DS Programs of the Faculty of Electronics, Telecommunications and Informatics, Gda\'nsk University of Technology, by the Swedish Research Council, Knut and by the Alice Wallenberg foundation. This work was also made possible through the support of grant from the John Templeton Foundation. The opinions expressed in this publication are those of the authors and do not necessarily reflect the views of the John Templeton Foundation. \\

\textbf{Author Contributions.}
R.R. initiated the research, proposed the Proposition 1 and provided the analytical proof of the Propositions 2 and 3. D.G. provided the analytic proof of the Proposition 1, and studied the feasibility of the experiment under presence of realistic errors. P.M. implemented numerical simulations. P. H. contributed to the development of the main idea and theoretical results. S. M and M. G designed and carried out the experiment. S.M. performed the data analysis. M. B. supervised the experimental part.  R.R. wrote the main text and  S.M. and M.B. the experimental part.  All the authors discussed the results and contributed to the final version of the manuscript. \\

\textbf{Competing interests.}
The authors declare no competing interests. 
\newpage

\renewcommand{\figurename}{Supplementary Figure}
\renewcommand{\tablename}{Supplementary Table}
\setcounter{figure}{0} 

\begin{center}
\large{\textbf{Supplementary Information: \\Steering is an essential feature of non-locality in quantum theory}}
\end{center}
\vspace{0.3cm}

\begin{center}
\textbf{Supplementary Note 1 - Formal Proofs}
\end{center}

\textit{Proposition 1.}
The optimal quantum strategy for the game $G^{(7)}$ (achieving $\omega_\text{q}(G^{(7)}) \approx 0.782$) violates the uncertainty principle - quantum game value correspondence, i.e., Alice is unable to steer Bob's system to the maximally certain states and vice versa.


	In the proof, we will use the Lemmas \ref{thm:Masanes} and \ref{lem:Jordan} stated below. 
	
	\begin{lemma}[\cite{MasanesSM}]
		\label{thm:Masanes}
		In the Bell scenario $B(n,2,2)$ for any number of parties $n$, all extreme boxes $P(a,b|x,y)$ of the quantum set $Q(n,2,2)$ can be realized by measuring $n$ qubit pure states with projective observables. 
	\end{lemma}
	\begin{lemma}[\textsc{Jordan's lemma} \cite{JordanSM}]
		\label{lem:Jordan}
		Any two binary observables $A_1$ and $A_2$ acting on a finite-dimensional Hilbert space $\mathbf{C}^n$ with $n \geq 1$ can be simultaneously block-diagonalized into $1 \times 1$ and $2 \times 2$ blocks.
	\end{lemma}
	Lemma \ref{thm:Masanes} allows us to find $\omega_\text{q}(G^{(7)})$ by optimizing over projective measurements on pure two qubit states. Let $(|\psi \rangle, \{M^{x}\}, \{M^{y}\})$ be a qubit strategy for the game $G^{(7)}$ with
	\be
	\label{eq:7-measurements}
	\ba
	M^{x} = \sum_{a = 0}^{1} (-1)^a \Pi^{x}_{a} = 
	\begin{pmatrix}
		0 & e^{i \alpha_x}  \\
		e^{ - i \alpha_x} & 0 
	\end{pmatrix}
	\ea
	\ee
	\be
	\ba
	M^{y} = \sum_{b = 0}^{1} (-1)^b \Pi^{y}_{b} = 
	\begin{pmatrix}
		0 & e^{i \beta_y}  \\
		e^{ - i \beta_y} & 0 
	\end{pmatrix},
	\ea
	\ee
	over the computational basis $\{ | 0 \rangle, |1 \rangle \}$, where without loss of generality $\alpha_0 = \beta_0 = 0$ and $\alpha_1, \beta_1 \in [- \pi, \pi]$. Here $\Pi^{x}_{a}$ and $\Pi^{y}_{b}$ are the projectors associated to the values $0, 1$ obtained by diagonalizing $M^{x}, M^{y}$ respectively. We can write the Bell operator in terms of the $\Pi^{x}_a$ and $\Pi^{y}_b$ as
	\be
	\mathbf{B}(G^{(7)}) = \!\sum_{x, y \in \{0,1\}} \pi_{AB}(x,y) \sum_{a,b \in \{0,1\}} V_{G^{(7)}}(a,b|x,y) \Pi^{x}_{a} \otimes \Pi^{y}_{b}, 
	\ee
	with $\pi_{AB}(x,y) = \pi_A(x) \pi_B(y)$ and $\pi_A(x) = \pi_B(y) = \frac{1}{2}$ for each $x,y \in \{0,1\}$. Our task is to find the optimal $\alpha_1, \beta_1$ that lead to the maximum eigenvalue of the operator $\mathbf{\tilde{B}}(G^{(7)}) = 4 \mathbf{B}(G^{(7)})$. The eigen-equation $\det[\mathbf{\tilde{B}}(G^{(7)})  - \lambda \mathbf{1} ] = 0$ simplifies to
	\be
	\label{eq:7-cycle-eigeneq}
	\ba
	\lambda & \left[-30 + \lambda (33 + 2 \lambda (\lambda -7) ) \right]  \\
	&  +  (\lambda -3) (\lambda -1) \left[ \cos{(\alpha_1)} - \cos{(\alpha_1)}  \cos{(\beta_1)} + \cos{(\beta_1)} \right]  \\
	& + 9  - \sin^2{(\alpha_1)} \sin^2{(\beta_1)} = 0.
	\ea
	\ee
	To find the optimum quantum value $\frac{1}{4} \lambda_{\mathrm{max}}$, we use the KKT conditions, i.e., we investigate the expression for $\lambda$ in terms of $\alpha_1$ and $\beta_1$ in four sectors.
	
	\underline{Case I}: $\alpha_1$ = $\beta_1$ = $\pi$. In this case, the eigen-equation (\ref{eq:7-cycle-eigeneq}) directly solves to $\lambda^{(I)} = 3$ corresponding to the (classical) game value of $\frac{3}{4}$.
	
	\underline{Case II}: $\beta_1$ = $\pi$, optimize over \textbf{$\alpha_1$.} In this case the eigen-equation simplifies to
	\be
	\label{eq:eigen-eq-simple}
	(\lambda^2 - 4 \lambda + 3) (1 - 3 \lambda + \lambda^2 + \cos{(\alpha_1)}) = 0,
	\ee
	which has the four solutions
	\be
	\lambda = 1,3, \frac{1}{2} \left(3 \pm \sqrt{5 - 4 \cos{\alpha_1}} \right),
	\ee
	so that we obtain the maximum value of $\lambda$ in this sector to be $3$, corresponding to a game value of $\frac{3}{4}$. 
	
	\underline{Case III}: $\alpha_1$ = $\pi$, optimize over \textbf{$\beta_1$.} As in the previous case II, the eigen-equation simplifies to Supplementary Equation (\ref{eq:eigen-eq-simple}) giving the maximum value of $\lambda = 3$ or a game value of $\frac{3}{4}$. 
	
	\underline{Case IV}: optimize over \textbf{$\alpha_1$, $\beta_1$.} Here we have $\frac{\partial \lambda}{\partial \alpha_1} = 0$ and $\frac{\partial \lambda}{\partial \beta_1} = 0$. We implicitly differentiate the eigen-equation (\ref{eq:7-cycle-eigeneq}) with respect to $\alpha_1$ and set $\frac{\partial \lambda}{\partial \alpha_1} = 0$ to get 
	\be 
	\label{eq:7-case4-1}
	(\lambda -3) (\lambda -1) \sin{(\alpha_1)} (-1 + \cos{(\alpha_1)}) = \sin{(2 \alpha_1)} \sin^2{(\beta_1)}.
	\ee
	Similarly implicit differentiation of Supplementary Equation (\ref{eq:7-cycle-eigeneq}) with respect to $\beta_1$ and setting $\frac{\partial \lambda}{\partial \beta_1} = 0$ gives
	\be 
	\label{eq:7-case4-2}
	(3 - \lambda) (\lambda -1) (1 - \cos{(\alpha_1)}) =  \sin^2{(\alpha_1)} \sin{(2 \beta_1)}.
	\ee
	Solving Eqs. (\ref{eq:7-case4-1}) and (\ref{eq:7-case4-2}) yields that either $\alpha_1 = 0$ giving $\lambda = 3$ or that the following relation holds between the optimal $\alpha_1$ and $\beta_1$
	\be
	\cos{(\alpha_1)} = \cos{(\beta_1)} \Longrightarrow \alpha_1 = \beta_1, 
	\ee
	since $\alpha_1, \beta_1 \in [-\pi, \pi]$. We can now simplify the eigen-equation (\ref{eq:7-cycle-eigeneq}) setting $\alpha_1 = \beta_1$. Implicit differentiation of the resulting expression with respect to $\alpha_1$ and again setting $\frac{\partial \lambda}{\partial \alpha_1} = 0$ gives the $\lambda$ in terms of the $\alpha_1$ as
	\be
	\label{eq:7-eig-alpha}
	\ba
	\left[ (-2 + \lambda)^2 + 2 \cos{(\alpha_1)} + \cos{(2 \alpha_1)} \right] \times \\
	\times \cos{\left(\frac{\alpha_1}{2} \right)} \sin^3{\left(\frac{\alpha_1}{2} \right)}  = 0, \\
	\Rightarrow \lambda = 2 \pm \sqrt{-2 \cos{(\alpha_1)} - \cos{(2 \alpha_1)} }.
	\ea
	\ee
	Substituting Supplementary Equation (\ref{eq:7-eig-alpha}) back into the eigen-equation 			given by Supplementary Equation (\ref{eq:7-cycle-eigeneq}) and solving gives the optimal 			$\alpha_1$
	\be
	\label{eq:7-alpha} 
	\alpha_1 =  2 \arctan{\sqrt{\frac{\eta^2 + 2 \eta - 5}{3 \eta} }},
	\ee
	with $\eta = \left[\frac{1}{2} (43 + 9 \sqrt{29}) \right]^{\frac{1}{3}}$. Thus the optimal value $\lambda_{\mathrm{max}}$ is given by Supplementary Equation (\ref{eq:7-eig-alpha}) with $\alpha_1$ in Supplementary Equation (\ref{eq:7-alpha}) as
	\be
	\ba
	\omega_\text{q}(G^{(7)}) & = \frac{\lambda_{\mathrm{max}}(\mathbf{\tilde{B}}(G^{(7)}))}{4} \\
	& = \frac{1}{108} (35 + (15740 - 972 \sqrt{29})^{\frac{1}{3}} \\
	& + 2^{\frac{2}{3}} (3935 + 243 \sqrt{29})^{\frac{1}{3}}) \\
	& \approx 0.7822.
	\ea
	\ee
	We thus have the optimal quantum strategy for game $G^{(7)}$ given by the measurement operators in Supplementary Equation (\ref{eq:7-measurements}) with $\alpha_1 = \beta_1$ given by Supplementary Equation (\ref{eq:7-alpha}). The corresponding optimal quantum (qubit) state is the maximal eigenvector of $\mathbf{B}(G^{(7)})$ with these measurements.
	
	
	Applying Jordan's Lemma \ref{lem:Jordan} to Alice's observables $M^{x}$ for $x = 0,1$, we obtain that
	\be
	M^{x} = \oplus_{i} M^{x}(i) = \sum_{i} M^{x}(i) \otimes |i \rangle \langle i|,
	\ee
	where $i$ labels the block index and each $M^{x}(i)$ is a $2 \times 2$ block (the $1 \times 1$ blocks can be enhanced by adding additional dimensions without loss of generality). This gives a basis in which the Hilbert space of Alice's observables $\mathcal{H}_{A}$ can be written as $\mathcal{H}_{A} = \mathbb{C}^{2} \otimes \mathcal{H'}_{A}$ with $\mathcal{H'}_{A}$ denoting the Hilbert space with basis $\{| i \rangle\}$ and $\mathbb{C}^{2}$ being a qubit space. A similar structure exists for the Hilbert space of Bob's observables $\mathcal{H}_{B}$. The crucial part of the above structure is that a measurement of the block index $i$ commutes with both Alice's observables $M^{x}$, so one can consider a general strategy in which Alice measures $| i \rangle \langle i|$ first, and similarly Bob measures the block index for his observables $M^{y}$. This reduces the whole problem to the case where $\mathcal{H}_{A} = \mathbb{C}^{2}$ and $\mathcal{H}_{B} = \mathbb{C}^{2}$. Within the qubit subspace, we know by Lemma \ref{thm:Masanes} that projective measurements on pure two qubit states achieve the extremal points, and as we have shown any qubit strategy is equivalent, up to local unitaries to the strategy given by the measurements in Supplementary Equation (\ref{eq:7-measurements}).
	
	Let us now examine the uncertainty relations corresponding to the game $G^{(7)}$ for each input-output pair $(x,a)$ of Alice. In what follows, we simplify notation by explicitly specifying numerical values to avoid cumbersome analytical expressions, however note that all the results are fully analytical. 
	\begin{eqnarray}
	&&	(x,a) = (0,0)\rightarrow \nonumber \\
	&& \qquad	P(b=0|y=0) + P(b=1|y=1) \leq 2 \xi_B^{(0,0)} \nonumber \\
	&&	(x,a) = (0,1) \rightarrow \nonumber \\
	&&	\qquad P(b=1|y=0) + P(b=0 | y = 1) \leq 2 \xi_B^{(0,1)} \nonumber \\
	&&	(x,a) = (1,0) \rightarrow \nonumber \\ 
	&&	\qquad P(b=1 | y = 0) +  P(b=1|y=1) \leq 2 \xi_B^{(1,0)} \nonumber \\
	&&	(x,a) = (1,1) \rightarrow P(b=0 |y=0) \leq 1,
	\end{eqnarray}
	where we have used $\pi_B(y=0) = \pi_B(y=1) = \frac{1}{2}$, and the uncertainty bounds are 
	\be
	\ba
	\xi_B^{(0,0)} = \xi_B^{(0,1)} = & \frac{1}{4} e^{-i \alpha_1} \left(2 e^{i \alpha_1} - \sqrt{-e^{i \alpha_1} (-1 + e^{i \alpha_1})^2} \right) \\
	\approx & 0.8815,
	\ea
	\ee
	and
	\be
	\xi_B^{(1,0)} = \frac{1}{4} e^{-i \frac{\alpha_1}{2}} \left(1 + e^{i \frac{\alpha_1}{2}} \right)^2 \approx 0.8232, \\
	\ee
	with the optimal $\alpha_1$ given in Supplementary Equation (\ref{eq:7-alpha}).

	\underline{Case I}: (x,a) = (0,0). The maximally certain state for $(x,a) = (0,0)$ is
	\be
	|\psi^{(0,0)} \rangle_A = \frac{1}{\sqrt{2}}  \left( \frac{\sqrt{-e^{i \alpha_1} (1 - e^{i \alpha_1})^2}}{ (1 - e^{i \alpha_1})} | 0 \rangle + |1 \rangle \right).
	\ee
	Projecting onto the optimal state with the optimal projector $\Pi^{x=0}_{a=0} = | + \rangle \langle + |$, we see that Alice only manages to steer Bob's system to the state 
	\be
	|\tilde{\psi} \rangle = \frac{1}{\sqrt{2}} \left( e^{- 1.146 i} | 0 \rangle + |1 \rangle \right),
	\ee
	which achieves value $0.8446 < \xi_B^{(0,0)}$ for the corresponding uncertainty relation. Note that here we have stated the expression numerically simply to avoid the cumbersome notation associated with the exact analytical expression.
	
	\underline{Case II}: (x,a) = (0,1). The maximally certain state for $(x,a) = (0,1)$ is 
	\be
	|\psi^{(0,1)} \rangle_A = \frac{1}{\sqrt{2}}  \left(  \frac{\sqrt{-e^{i \alpha_1} (1 - e^{i \alpha_1})^2}}{ (-1 + e^{i \alpha_1})} | 0 \rangle + | 1 \rangle \right).
	\ee
	Projecting onto the optimal state with the optimal projector $\Pi^{x=0}_{a=1} = | - \rangle \langle - |$, we see that Alice only manages to steer Bob's system to the state 
	\be
	|\tilde{\psi} \rangle = \frac{1}{\sqrt{2}}  \left( -e^{- 0.2598 i} | 0 \rangle + | 1 \rangle \right),
	\ee
	which achieves value $0.8446 < \xi_B^{(0,1)}$ for the corresponding uncertainty relation. \\
	
	\underline{Case III}: (x,a) = (1,0). The maximally certain state for $(x,a) = (1,0)$ is
	\be
	\label{eq:7-uncert1}  
	|\psi^{(1,0)}\rangle_A = \frac{1}{\sqrt{2}} \left(  - e^{i \frac{\alpha_1}{2}} | 0\rangle + |1 \rangle \right).
	\ee
	Projecting onto the optimal state with the projector $\Pi^{x=1}_{a=0}$, we see that Alice manages to steer Bob's system to the maximally certain state in Supplementary Equation (\ref{eq:7-uncert1}) for this uncertainty relation.
	
	\underline{Case IV}: (x,a) = (1,1). The maximally certain state for $(x,a) = (1,1)$ corresponds to the projector $\Pi^{y=0}_{b=0} = | + \rangle \langle + |$. Projecting onto the optimal state with the projector $\Pi^{x=1}_{a=1}$, we see that Alice only manages to steer Bob's system to the state 
	\be
	|\tilde{\psi} \rangle = \frac{1}{\sqrt{2}} \left( e^{0.36 i} | 0 \rangle + |1 \rangle \right),
	\ee
	which achieves value $0.968 < \xi_B^{(1,1)} = 1$ for this uncertainty relation.
	
	We thus see that for the game $G^{(7)}$, Alice is unable to steer Bob's system to the maximally certain states even for the non-trivial uncertainty relations corresponding to $(x,a) = (0,0)$ and $(x,a) = (0,1)$. In order to formally complete the argument for every optimal quantum strategy, we note that in the general case, one obtains a mixture of uncertainty relations over the outcomes $i$ of the block index measurement by Alice and Bob. Since the uncertainty relation fails to be saturated in each block, this implies that the same holds true also in the convex mixture of uncertainty relations.

	The quantum value of the game $\omega_\text{q}(G^{(7)})$ is thus lower than what could have been achieved if the non-locality of the theory were bounded by the uncertainty principle alone. The same holds true from the point of view of Bob steering Alice's system. While Bob is able to steer Alice's system to the maximally certain state for $(y,b) = (1,1)$, he is unable to do so for the non-trivial uncertainty relations corresponding to $(y,b) = (0,0)$ and $(y,b) = (0,1)$ as well as for the trivial uncertainty relation for $(y,b) = (1,0)$. Thus, this example proves that the non-locality of quantum theory is not determined by the uncertainty principle alone, and steering plays a definite role. 

\begin{center}
\textbf{Supplementary Note 2 - Relation with the Schrodinger-Hughston-Jozsa-Wootters theorem}
\end{center}

Let us remark upon a curious feature of the rewriting in Eq.(2) of the main text 
with regard to steerability of quantum systems. Consider a set of measurement operators $M^{x}_{a}$ on Alice's side, i.e., positive operators $M^{x}_{a} \geq 0$ satisfying $\sum_{a} M^{x}_{a} = \mathbf{1}$. Such a collection represents a positive-operator valued measure (POVM) for each $x$. For any fixed bipartite quantum state $\hat{\sigma}_{AB}$, every measurement on Alice's side gives rise to an assemblage $\{\sigma^B_{a|x}\}_{a,x} = \{P_\text{A$|$X}(a|x), \hat{\sigma}^\text{B}_{a|x}\}_{a,x}$. Here
\be 
\label{eq:assemblage-quant}
\sigma^\text{B}_{a|x} = \tr_{A} \left[(M^{x}_{a} \otimes \mathbf{1}) \hat{\sigma}_{AB} \right],
\ee  
are the conditional (unnormalised) states of Bob's system prepared by Alice's measurement. We have that 
\be 
\label{eq:steering-ns}
\sum_{a} P_\text{A$|$X}(a|x) \hat{\sigma}^\text{B}_{a|x} = \sum_{a} P_\text{A$|$X}(a|x') \hat{\sigma}^\text{B}_{a|x'} = \hat{\sigma}_{B} = \tr_{A} \hat{\sigma}_{AB},
\ee 
for every $x,x'\in\mathcal{X}$, in order to obey the no-signaling principle; i.e., without the knowledge of Alice's outcome $a$, Bob's state is independent of the measurement choice $x$. The well-known Schr\"{o}dinger-Hughston-Jozsa-Wootters (SHJW) theorem \cite{HJWSM, SchrodingerSM} 
shows that every assemblage $\{P_\text{A$|$X}(a|x), \hat{\sigma}^\text{B}_{a|x}\}_{a,x}$ satisfying Supplementary Equation (\ref{eq:steering-ns}) has a quantum realization as in Supplementary Equation (\ref{eq:assemblage-quant}) for some quantum state $\hat{\sigma}_{AB}$ and for some set of measurement operators $\{M^{x}_{a}\}$. Now, 
the set of states $\{\tilde{\sigma}^\text{B}_{a|x}\}$ achieving the maximum value of the uncertainty relations 
together with the optimal probabilities $\{P_{A|x}(a|x)\}$ forms an assemblage. 
One might then wonder whether the result of \cite{OWSM} is a direct consequence of the SHJW theorem, since Alice might steer to the assemblage corresponding to the maximally certain states. However, maximally certain states together with the optimal local probabilities $P(a|x)$ do not guarantee that the no-signaling principle (\ref{eq:steering-ns}) holds. Thus, the UP-QGV correspondence found in \cite{OWSM} is a non-trivial property of the optimal states and measurements.
It was posed as an open question in \cite{OWSM} whether the correspondence holds for all non-local games. 

\begin{center}
\textbf{Supplementary Note 3 - Games vs. General Bell inequalities}
\end{center}

As observed in the main text, it is crucial for the correspondence between uncertainty relations and optimal quantum strategy to be meaningful that the form of the Bell expressions is restricted, for instance to the form of non-local games as in the paper \cite{OWSM}. It is readily seen that if one allows an arbitrary freedom in rewriting the Bell expressions up to normalization and no-signaling equality constraints  as suggested in \cite{ScaraniSM}, then one can always find a form of the Bell expression where the correspondence holds approximately, up to an arbitrarily small error. For instance, consider the following Bell expression, where the game has been supplemented with multiples $\lambda_{x,y}$ of the normalization constraints for input pairs $(x,y) \in \mathcal{X} \times \mathcal{Y}$
\ben 
&&\sum_{x,y} \pi_{A,B}(x,y) \sum_{a,b} V(a,b,x,y) P_\text{A,B$|$X,Y}(a,b|x,y)  \nonumber \\
&& \qquad \qquad + \sum_{x,y} \lambda_{x,y} \sum_{a,b} P_\text{A,B$|$X,Y}(a,b|x,y) \leq \beta_c. 
\een
When $\lambda_{x,y}$ factorize as $\lambda_{x,y} = \pi_{A}(x) \beta_y$, the resulting fine-grained uncertainty relation for fixed $(x,a)$ is given as
\ben
\sum_{y,b} \pi_{B}(y|x) V(a,b,x,y) P_\text{B$|$Y,X,A}(b|y,x,a) + \sum_{y} \beta_y \leq \tilde{\xi}_{x,a}, \nonumber \\
\een
where we have used $\sum_{b} P_\text{B$|$Y,X,A}(b|y,x,a) = 1$ for all $y$ in the second term. Now clearly, the absolute value of the multipliers $\beta_y$ need to be bounded at least as $\vert \beta_y \vert \leq 1$ to be comparable with $\pi_{B}(y|x)$. Similar considerations hold also for the multipliers associated to the no-signaling constraints. Failing such restrictions, one might always choose appropriately large (in comparison with $\pi_{B}(y|x)$) $\beta_y$ that lead to the saturation of $\tilde{\xi}_{x,a}$ in the uncertainty relation, up to an arbitrary small deviation. Even otherwise, the artificial addition of normalization and no-signaling constraints which are satisfied by all boxes in the set, leads to the question whether the resulting saturation of the uncertainty relations is intrinsic to the non-local correlations that maximally violate the inequality. To avoid such mathematical sleight of hand (which is also inherent in questions such as that of unbounded violation of Bell inequalities \cite{UnbSM}, the unique games conjecture \cite{UGCSM}, etc.) we follow \cite{OWSM} in restricting to non-local games, i.e., where $P_\text{A,B$|$X,Y}(a,b|x,y)$ appear in the Bell expression only with non-negative coefficients and all non-zero coefficients are equal (to $\pi_{A,B}(x,y)$) for a fixed input pair $(x,y)$. Note that this difference between non-local games and general Bell inequalities has also been noted previously in \cite{SMA07SM}.

\begin{center}
\textbf{Supplementary Note 4 - Constructing counter-example games for all non-maximally entangled states}
\end{center}

In the previous sections, we have seen an example of a non-local game with the optimal state being a non-maximally entangled two-qubit state, for which the UP-QGV correspondence breaks down. In this section, we prove that this is not a one-off instance, indeed for every non-maximally entangled two-qudit state $| \psi \rangle \in \mathbb{C}^d \otimes \mathbb{C}^d$ (for an Hilbert space of arbitrary dimension $d$), one can construct a two-player game $G_{\psi}$ such that $| \psi \rangle$ is optimal for $G_{\psi}$ and such that the UP-QGV correspondence does not hold for $G_{\psi}$. We present the construction in this section as the proof of the following Proposition from the main text. The construction we use resembles that used by Coladangelo et al. in \cite{CGS16} to show that all bipartite pure entangled states can be self-tested.

\textit{Proposition 3.}
For any two-party entangled, but non-maximally entangled, state $| \psi \rangle \in \mathbb{C}^d \otimes \mathbb{C}^d$ for arbitrary Hilbert space dimension $d$, there exists a game $G_{\psi}$ for which the optimal quantum strategy is given by suitable measurements on $| \psi \rangle$, and such that the correspondence between the uncertainty principle and the quantum game value does not hold for $G_{\psi}$.

\begin{proof}
	The first step in the construction is to show the statement for all entangled, but non-maximally entangled, two-qubit states; i.e., states of the form
	\begin{equation}
	| \psi_{\theta} \rangle = \cos{\theta} |00 \rangle + \sin{\theta} |11 \rangle,
	\end{equation}
	with $\theta \in (0,\frac{\pi}{4})$. To do this, we use a slightly different game than $G^{(7)}$, namely the tilted CHSH inequality of \cite{tiltCHSHSM}, which we reformulate as a game. In the tilted CHSH game $\text{CHSH}_{\text{tilt}}$, Alice and Bob each receive inputs $x \in \{0,1\}$ and $y \in \{0,1,2\}$ with probabilities $\pi_{X}(0) = \frac{2+\beta}{4+\beta}$, $\pi_{X}(1) = \frac{2}{4+\beta}$, and $\pi_\text{Y$|$X}(0|0) = \pi_\text{Y$|$X}(1|0) = \frac{1}{2+\beta}$, $\pi_\text{Y$|$X}(2|0) = \frac{\beta}{2+\beta}$, $\pi_\text{Y$|$X}(0|1) = \pi_\text{Y$|$X}(1|1) = \frac{1}{2}$, $\pi_\text{Y$|$X}(2|1) = 0$, with $\beta = \frac{2}{\sqrt{1+2 \tan^2{2\theta}}}$, and return binary answers $a,b \in \{0,1\}$. The winning constraints for the game are the same as the usual CHSH game for $x,y \in \{0,1\}^2$, i.e., $V(a,b|x,y) = 1$ if $a \oplus b = x \cdot y$, while for $x=0,y=2$, $V(a,b|x,y) = 1$ if $a=0$. The Bell expression for the game is thus given by
	\begin{eqnarray}
	\frac{1}{4 + \beta} \sum_{x,y=0,1} P(a \oplus b = x \cdot y|x,y) +&&\frac{\beta}{4 + \beta} P(a=0|0,2) \nonumber \\ && \qquad  \leq \frac{3 + \beta}{4+ \beta}.
	\end{eqnarray}
	The classical value of the game is $\omega_c(\text{CHSH}_{\text{tilt}}) = \frac{3 + \beta}{4+\beta}$. The quantum value of the game is $\omega_\text{q}(\text{CHSH}_{\text{tilt}}) = \frac{1}{2} + \frac{\sqrt{8+2 \beta^2}}{8+2\beta}$ and is achieved when the parties perform the following measurements $A_x, B_y$ on the state $| \psi_{\theta} \rangle$:
	\begin{eqnarray}
	&&A_0 = \sigma_z, \qquad A_1 = \sigma_x, \nonumber \\
	&&B_0 = \cos{\mu} \sigma_z + \sin{\mu} \sigma_x, B_1 = \cos{\mu} \sigma_z - \sin{\mu} \sigma_x, B_2 = \sigma_z, \nonumber \\
	\end{eqnarray} 
	with $\mu = \arctan{(\sin{2 \theta})}$.
	
	The corresponding uncertainty relations are given as
	\begin{eqnarray}
	&&	(x,a) = (0,0)\rightarrow \nonumber \\
	&& P_\text{B$|$Y,X,A}(0|0,(0,0)) + P_\text{B$|$Y,X,A}(0|1,(0,0)) + \nonumber \\&& \beta P_\text{B$|$Y,X,A}(0|2,(0,0)) + \beta P_\text{B$|$Y,X,A}(1|2,(0,0)) \leq (2+\beta) \xi_B^{(0,0)} \nonumber \\
	&&	(x,a) = (0,1) \rightarrow \nonumber \\
	&& P_\text{B$|$Y,X,A}(1|0,(0,1)) + P_\text{B$|$Y,X,A}(1|1,(0,1)) \leq (2+\beta) \xi_B^{(0,1)} \nonumber \\
	&&	(x,a) = (1,0) \rightarrow \nonumber \\ 
	&& P_\text{B$|$Y,X,A}(0|0,(1,0)) +  P_\text{B$|$Y,X,A}(1|1,(1,0)) \leq 2 \xi_B^{(1,0)} \nonumber \\
	&&	(x,a) = (1,1) \rightarrow \nonumber \\
	&& P_\text{B$|$Y,X,A}(1|0,(1,1)) + P_\text{B$|$Y,X,A}(0|1,(1,1)) \leq 2 \xi_B^{(1,1)}, \nonumber \\
	\end{eqnarray}
	where $\xi_B^{(x,a)}$ are the bounds on the uncertainty expressions. We find that while the first two inequalities above are saturated by the optimal quantum strategy, the third and fourth inequalities fail to be saturated except when $\theta = \frac{\pi}{4}$, i.e., for the maximally entangled state $| \psi_{\frac{\pi}{4}} \rangle$. For the third expression, i.e., when $(x,a) = (1,0)$, the bound is
	\begin{eqnarray}
	\xi_B^{(1,0)} = \frac{\sqrt{3-\cos{4\theta}} + \sqrt{2} \sin{2\theta}}{2 \sqrt{3-\cos{4\theta}}}
	\end{eqnarray}
	with the maximally certain state being $| + \rangle = \frac{1}{\sqrt{2}}(|0 \rangle + |1 \rangle)$. On the other hand, we see that for $(x,a) = (1,0)$ using the optimal strategy Alice steers Bob's state to $| \tilde{\psi}^{(1,0)} \rangle = \cos{\theta} |0 \rangle + \sin{\theta} |1 \rangle$, so that only for the maximally entangled state ($\theta = \frac{\pi}{4}, \beta = 0$) does Alice manage to steer Bob's system to the least uncertain state. Similarly for the case $(x,a) =(1,1)$, the bound is 
	\begin{eqnarray}
	\xi_B^{(1,1)} = \frac{\sqrt{3-\cos{4\theta}} + \sqrt{2} \sin{2\theta}}{2 \sqrt{3-\cos{4\theta}}}
	\end{eqnarray}
	with the maximally certain state being $| - \rangle = \frac{1}{\sqrt{2}}(|0 \rangle - |1 \rangle)$. On the other hand, for $(x,a) = (1,1)$ using the optimal strategy Alice steers Bob's state to $|\tilde{\psi}^{(1,1)} \rangle = \cos{\theta} |0 \rangle - \sin{\theta} |1 \rangle$. So that it is again only for the maximally entangled state that Alice manages to steer Bob's system to the least uncertain state. Thus, the tilted CHSH inequality of \cite{tiltCHSHSM} expressed as a game, shows that every non-maximally entangled two-qubit state serves as the optimal state for a game in which the uncertainty principle - quantum game value correspondence does not hold.
	
	It now remains to generalize the result even further, to all two-qudit states that are non-maximally entangled. 
	Consider a general two-qudit non-maximally entangled state, written as
	\begin{eqnarray}
	| \phi \rangle = \sum_{i=1}^{d} \lambda_i |i, i \rangle
	\end{eqnarray} 
	with the Schmidt coefficients $\lambda_i \in \mathbb{R}$ obeying $0 < \lambda_i < 1$ for all $i$ and $\sum_i \lambda_i^2 = 1$ with not all $\lambda_i$ equal to $\frac{1}{\sqrt{d}}$. We first deal with the case when $d$ is even. The idea is to design a game with $d$-outcome measurements on each side, such that the correlation tables for some measurement settings are block-diagonal with blocks of size $2 \times 2$ each. The $j$-th $2 \times 2$ block will correspond to a tilted CHSH game that is maximally violated by a two qubit state, which is a normalized version of $\lambda_{2j-1} |2j-1, 2j-1 \rangle + \lambda_{2j} |2j, 2j \rangle$, with $j = 1, \dots, d/2$. Accordingly, we construct a game with two inputs $x = 0,1$ for Alice, and $2d+2$ inputs $y = 0, \dots, 2d+1$ for Bob, with $d$ outputs each. Given $| \phi \rangle$, the ratios $\left\{ \frac{\lambda_{2j}}{\lambda_{2j-1}} \right\}$ for $j = 1, \dots, d/2$ determine the game as follows. For inputs $x=0,1$ and $y=0, \dots, d/2+1$, the players play a set of $d/2$ tilted CHSH games determined by the following procedure. 
	
	Set $\theta_{2j} = \arctan{\frac{\lambda_{2j}}{\lambda_{2j-1}}}$, with corresponding optimal states $|\psi_{\theta_{2j}} \rangle = \cos{\theta_{2j}} | 00 \rangle + \sin{\theta_{2j}} |11 \rangle$. We obtain the set of $d/2$ tilted CHSH games with tilting parameters $\beta_{2j}$ for $j =1, \dots, d/2$ given by
	\begin{eqnarray}
	\beta_{2j} = \frac{2}{\sqrt{1+2 \tan^2{2 \theta_{2j}}}}.
	\end{eqnarray} 
	The $j$-th tilted CHSH game $\text{CHSH}^{(j)}_{\text{tilt}}$ is played on inputs $x=0,1$ for Alice and $y = 0, 1, j+1$ for Bob, so that each of the settings $y=2, \dots, d+1$ appears in a distinct tilted CHSH game. With an appropriate choice of observables, the players are able to achieve a value proportional to $(\lambda_{2j}^2+\lambda_{2j-1}^2) \omega_\text{q}^{(2j)}$ for the game with \begin{eqnarray}
	\omega_\text{q}^{(2j)} := \left(\frac{1}{2} + \frac{\sqrt{8+2 \beta_{2j}^2}}{8+2 \beta_{2j}}\right) (4+\beta_{2j})
	\end{eqnarray} 
	for $j=1,\dots, d/2$. Let $\omega_\text{q}^* := \max_j \omega_\text{q}^{(2j)}$, and note that $\omega_\text{q}^*$ is determined once the $\lambda_i$ are given.
	
	To complete the construction, we now consider another set of $d/2$ tilted CHSH games with parameters $\beta_{2j}$, this time played by Alice and Bob on the inputs $x=0,1$ and $y = d/2+3j-1, d/2+3j, d/2+3j+1$ for $j=1, \dots, d/2$.
	It remains to specify the input distributions, these are given with $\tau := 4 + \sum_{j=1}^{d/2} \left( \beta_{2j} + \frac{\omega_\text{q}^*- \omega_\text{q}^{(2j)}}{\omega_\text{q}^{(2j)}}(4+\beta_{2j}) \right)$ as
	\begin{eqnarray}
	&&\pi_X(0) = \frac{1}{\tau} \left[2 + \sum_j \left(\beta_{2j} + (2+\beta_{2j})\frac{\omega_\text{q}^* - \omega_\text{q}^{(2j)}}{\omega_\text{q}^{(2j)}}  \right) \right], \nonumber \\
	&&\pi_{X}(1) = 1 - \pi_{X}(0) = \frac{2+2\sum_{j=1}^{d/2}(\omega_\text{q}^*-\omega_\text{q}^{(2j)})/(\omega_\text{q}^{(2j)})}{\tau}, \nonumber \\
	&&\pi_\text{Y$|$X}(0|0) = \pi_\text{Y$|$X}(1|0) = \frac{1}{\pi_{X}(0) \tau}, \nonumber\\
	&&\pi_\text{Y$|$X}(j+1|0) = \frac{\beta_{2j}}{\pi_{X}(0) \tau}, \quad j = 1, \dots, d/2, \nonumber \\
	&&\pi_\text{Y$|$X}(d/2+3j-1|0) = \pi_\text{Y$|$X}(d/2+3j|0) = \frac{\omega_\text{q}^* - \omega_\text{q}^{(2j)}}{\omega_\text{q}^{(2j)} \pi_X(0) \tau}, \nonumber \\
	&&\pi_\text{Y$|$X}(d/2+ 3j+1|0) = \frac{\beta_{2j}(\omega_\text{q}^* - \omega_\text{q}^{(2j)})}{\omega_\text{q}^{(2j)} \pi_X(0)\tau}, \nonumber \\
	&&\pi_\text{Y$|$X}(d/2+3j-1|1) = \pi_\text{Y$|$X}(d/2+3j|1) = \frac{\omega_\text{q}^* - \omega_\text{q}^{(2j)}}{\omega_\text{q}^{(2j)} \pi_X(1) \tau}, \nonumber \\
	&&\pi_\text{Y$|$X}(0|1) = \pi_\text{Y$|$X}(1|1) = \frac{1}{\pi_X(1) \tau}.
	\end{eqnarray}
	With the above input distributions, we can now directly calculate the value achieved by a quantum strategy given by the shared state $| \phi \rangle$ and observables $A_0 = \oplus_{j=1}^{d/2} \sigma_z^{(j)}$, $A_1 = \oplus_{j=1}^{d/2} \sigma_x^{(j)}$, $B_0 = B_{d/2+3j-1} = \oplus_{j=1}^{d/2} \left(\cos{\mu} \sigma_z^{(j)} + \sin{\mu} \sigma_x^{(j)} \right)$ and $B_1 = B_{d/2+3j} = \oplus_{j=1}^{d/2} \left( \cos{\mu} \sigma_z^{(j)} - \sin{\mu} \sigma_x^{(j)} \right)$, $B_{j+1} = B_{d/2+3j+1} = \oplus_{j=1}^{d/2} \sigma_z^{(j)}$. In the $j$-th $2 \times 2$ sector, the strategy achieves a value of $(\lambda_{2j}^2 + \lambda_{2j-1}^2)  \frac{\omega_\text{q}^*}{\tau}$, so that summing over all $j = 1, \dots, d/2$, we obtain the quantum value of our generalized tilted CHSH game to be 
	\begin{eqnarray}
	\omega_\text{q}(\text{CHSH}_{\text{gen-tilt}}) = \sum_{j=1}^{d/2} (\lambda_{2j}^2 + \lambda_{2j-1}^2)  \frac{\omega_\text{q}^*}{\tau} = \frac{\omega_\text{q}^*}{\tau},
	\end{eqnarray}
	by virtue of the fact that $\sum_i \lambda_i^2 = 1$. Let us now verify that this is in fact the optimal quantum value of the game $\text{CHSH}_{\text{gen-tilt}}$. This is seen by the fact that the game decomposes into $2 \times 2$ blocks, and the maximum quantum value within each block is $\frac{\omega_\text{q}^*}{\tau}$, obtained from $\omega_\text{q}(\text{CHSH}_{\text{tilt}})$ presented earlier. Moreover, we see that the uncertainty relations fail to be saturated within each $2 \times 2$ block, except those which correspond to $\lambda_{2j} = \lambda_{2j-1}$, from the results for the qubit case. Finally, the case for odd $d$ works in a very similar manner to that for even $d$, we use the generalized tilted CHSH game corresponding to $d-1$ which is even, and augment the game with a $1 \times 1$ block, i.e., the entries $P_\text{A,B$|$X,Y}(d,d|x,y)$ for $x,y \in \{0,1\}^2$ and $x=0,1$, $y=2d, 2d+1$. Similarly, we augment the observables with the projector $|d \rangle \langle d|$, i.e., $A_0 = \oplus_{j=1}^{(d-1)/2} \left(\sigma_z^{(j)} \oplus |d \rangle \langle d| \right)$. While the uncertainty relation corresponding to the $1 \times 1$ block is saturated, for all non-maximally entangled states, the uncertainty relations for the $2 \times 2$ blocks are not, as we have seen in the even $d$ case. Thus, we have constructed for every $d \geq 2$, a non-local game with the optimal strategy being given by the state $| \phi \rangle = \sum_i \lambda_i | i, i \rangle$ and such that the correspondence between the uncertainty principle and the quantum game value is broken. 
\end{proof}

\begin{center}
\textbf{Supplementary Note 5 - Experimental implementation}
\end{center}

In the experiment, we used single photon's polarization state as the physical qubit. To maximally violate the Bell inequality given in Eq. $(5)$ of the main text, we prepare the following polarization entangled two-photons state,
\begin{eqnarray}{\label{eqE_1}}
|\Psi\rangle =& 0.2487 |HH\rangle &+ \hspace{2mm} 0.4760 |HV\rangle  \\ \nonumber
+&  0.8060 |VH\rangle &- \hspace{2mm} 0.2487 |VV\rangle.
\end{eqnarray}
This state is produced as follows, firstly, we prepare maximally entangled photons pairs (see Fig. $(3)$ of the main paper). For this an Ultraviolet light centered at wavelength of 390\,nm was focused onto two 2\,mm thick $\beta$ barium borate (BBO) nonlinear crystals placed in cross-configuration to produce photon pairs emitted into two spatial modes $a$ and $b$ through the second order degenerate type-I spontaneous parametric down-conversion (SPDC) process \cite{SPDC_crossSM}. Any spatial or temporal distinguishability between the down-converted photons is carefully removed through quartz wedges placed in the pump beam (not shown in the figure) and a pair of $YVO_4$ crystals located in each of the down-converted beams. The emitted photons were passed through the narrow-bandwidth interference filters (IF) ($\Delta\lambda=3$\,nm) and coupled into 2\,m single mode optical fibers (SMF) to secure well defined spatial and spectral emission modes. Secondly, to prepare the desired state as outlined
in \cite{State_prepSM},  the pump polarization is altered to produce the state $\cos{\theta_p}|HH\rangle - \sin{\theta_p}|VV\rangle$ with $\theta_p =31.50^{\circ}$. Then, as the final step this state is rotated to $|\Psi\rangle$ by the use of a half wave-plate (HWP) placed after the output fiber coupler in each of the mode (a) and (b) at an angle of $39.69^\circ$ and $84.69^\circ$ respectively.

\begin{center}
\textbf{Supplementary Note 6 - State tomography}
\end{center}

To estimate the fidelity of the two-photon prepared state $(\rho_{\text{exp}})$ with respect to the $\rho_{\text{th}}=|\Psi\rangle\langle\Psi|$, we carried out quantum state tomography as described in Ref. \cite{tomoSM}, and where we have measured each of the two photons in three mutually unbiased basis (H/V, D/A, L/R). These polarization measurements were performed by using HWPs, quarter wave plates (QWP) and polarizing beam splitters (PBS) followed by single photon detectors (actively quenched Si-avalanche photodiodes (Si-APD)). An FPGA based timing system is used to record the number of coincidence events with a detection time window of $1.7$~ns. For each setting, we have obtained approximately 1.6 Million events.

The obtained density matrix of the prepared state is 	

\begin{widetext}
	\begin{equation}
	\rho_{exp} =\left(
	\begin{array}{cccc}
	0.065 & 0.124\, -0.001 i & 0.197\, -0.013 i & -0.059-0.002 i \\
	0.124\, +0.001 i & 0.242 & 0.39\, -0.025 i & -0.116-0.005 i \\
	0.197\, +0.013 i & 0.39\, +0.025 i & 0.638 & -0.186-0.021 i \\
	-0.059+0.002 i & -0.116+0.005 i & -0.186+0.021 i & 0.056 \\
	\end{array}
	\right),
	\end{equation}
\end{widetext}
where entry--wise error bars are given by
\begin{equation}
\Delta\rho_{exp}= \left(
\begin{array}{cccc}
0.002 & 0.004 & 0.007 & 0.003 \\
0.004 & 0.001 & 0.010 & 0.004 \\
0.007 & 0.010 & 0.002 & 0.007 \\
0.003 & 0.004 & 0.007 & 0.002 \\
\end{array}
\right).
\end{equation}
The real and imaginary parts of this density matrix are shown in Supplementary figure \ref{fig3}. From experimentally collected data, we reconstructed the density matrix $\rho_{exp}$ by considering maximum likelihood estimation, as described in Ref. \cite{tomoSM}. Our figure of merit to quantify effectiveness of state reconstruction is the relative fidelity with respect to the theoretical state $|\Psi\rangle$. Errors for fidelity reconstruction have been estimated by taking into account the Poissonian statistical distribution of the photon number counting (see Section VI, Ref. \cite{tomoSM}). 
\begin{figure}[ht!]
        \begin{center}
                \centerline{\includegraphics[width=1\columnwidth]{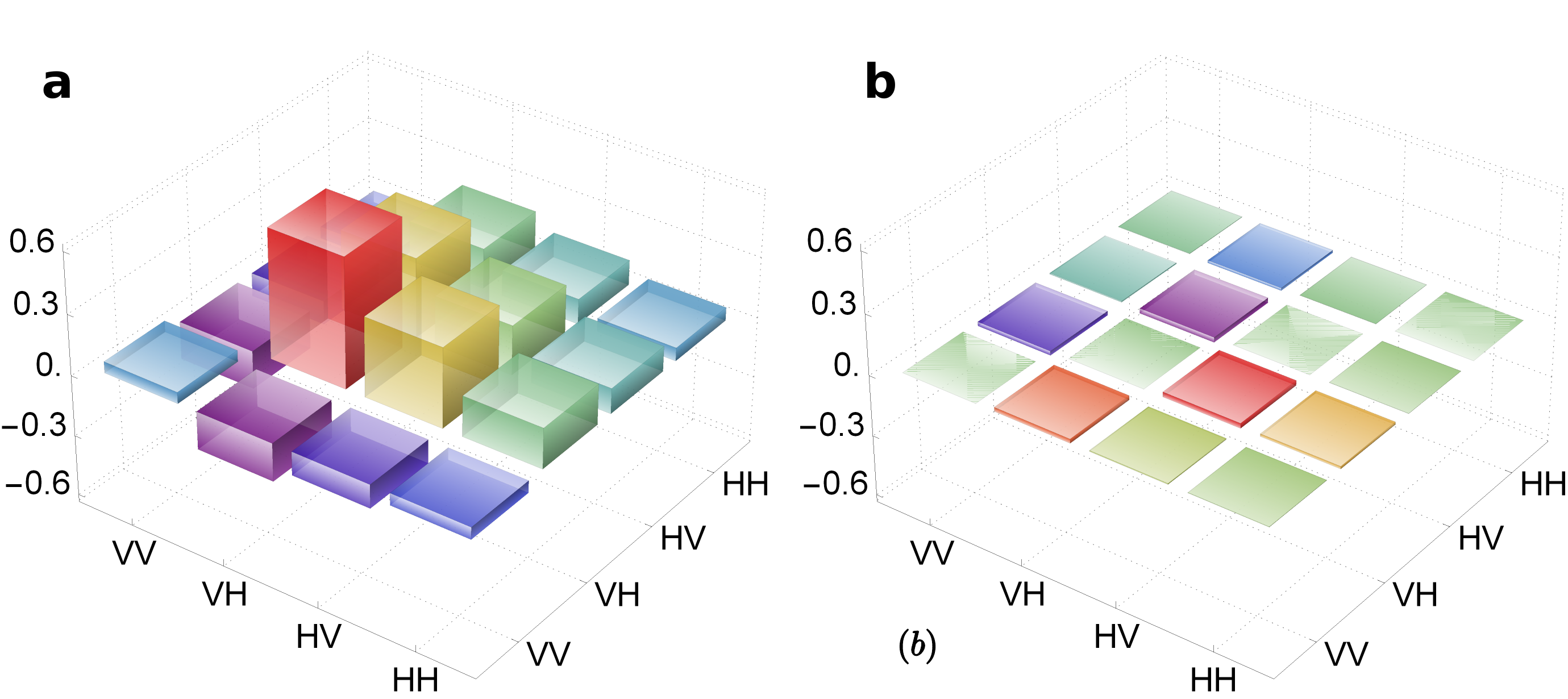}}
                \caption[Experimental density matrix.]{(Color online). Experimental density matrix. \textbf{a} Real and, \textbf{b} imaginary part, shown in computational basis (HH, HV,VH and VV), of the experimentally obtained elements of the two qubit density matrix  $\rho_{Exp}$, constructed using maximum likelihood quantum state tomography.}
                \label{fig3}
        \end{center}
\end{figure}
We obtained the following state fidelity 
\begin{equation}
F= \langle\Psi|\rho_{exp}|\Psi\rangle = 0.9933 \pm 0.0009 .
\end{equation}

\begin{center}
\textbf{Supplementary Note 7 - Steering and Bell Violation}
\end{center}
	
Alice and Bob can perform Bell test on experimentally prepared state with the following settings \\
\begin{equation}
\begin{split}
|\phi_0^{\pm}\rangle =& \frac{1}{\sqrt{2}}\left(|H\rangle \pm  |V\rangle \right) \\
|\phi_1^{\pm}\rangle =& \cos \gamma  |H\rangle \pm \sin \gamma  |V\rangle
\end{split}
\end{equation}
where $\gamma=4.794814$. 

To perform Bell test and the tomography of the steered state, Bob randomly chooses if he wants to realize tomographic measurement or the Bell test settings. Sequence of measurements performed for each task are given in the table \ref{Table_1}. These choices are randomly made and executed via computer controlled rotation stages carrying wave-plates of Bob's analyzer. To decrease the statistical counts error further, we have collected approximately 6.48 Million events per setting.
\begin{table}[t]
	\centering
	\caption{\label{Table_1} For a particular setting of Alice (column 1), Bob chooses randomly if he wants to perform sequence of measurements for the Bell test (column 2, upper row for the corresponding Alice setting) or tomography of the photon he possess (column 2, lower row for the corresponding Alice setting).}
	{\begin{tabular}{ |l|l| }
			\hline
			{\bf Alice Settings} & {\bf Bob Settings} \\ \hline \hline
			\multirow{2}{*}{\hspace{7mm} \vspace{1mm}$|\phi_1^{\pm}\rangle$ }
			& \hspace{4mm} $|\phi_1^{\pm}\rangle$ ,  $|\phi_2^{\pm}\rangle$  \\	\cline{2-2}
			& H/V, D/A, L/R \\ \hline
			\multirow{2}{*}{\hspace{7mm} \vspace{1mm}$|\phi_2^{\pm}\rangle$ }
			& \hspace{4mm} $|\phi_1^{\pm}\rangle$ ,  $|\phi_2^{\pm}\rangle$  \\	\cline{2-2}
			& H/V, D/A, L/R \\ \hline
			
			\hline
	\end{tabular} }
\end{table}

For the Bell test we obtained $\omega_\text{q}(G^{(7)})= 0.7770\pm 0.0002$, when the corresponding theoretical value is  $0.78221$. In the following, we report theoretical and the experimentally obtained density matrices of  Bob states when Alice projected her photon. 
\begin{enumerate}
	
	\item When Alice's photon is projected to $|\phi_0^{+}\rangle$, Bob state becomes \\ \\
	$\left(
	\begin{array}{cc}
	0.9556 & 0.2060 \\
	0.2060 & 0.0444 \\
	\end{array}
	\right)$ \\
	
	and we  obtained experimentally \\
	
	$\left(
	\begin{array}{cc}
	0.943 & 0.231\, -0.009 i \\
	0.231\, +0.009 i & 0.057 \\
	\end{array}
	\right)$\\
	$\pm \left(
	\begin{array}{cc}
	0.002 & 0.011 \\
	0.011 & 0.002 \\
	\end{array}
	\right)$, 	which corresponds to the fidelity of $ 0.9990 \pm 0.0003$.
	
	\item When Alice photon is projected to $|\phi_0^{-}\rangle$, Bob state becomes \\
	
	$\left(
	\begin{array}{cc}
	0.3716 & -0.4832 \\
	-0.4832 & 0.6285 \\
	\end{array}
	\right)$ \\
	
	and we  obtained experimentally \\
	
	$\left(
	\begin{array}{cc}
	0.369 & -0.471-0.051 i \\
	-0.471+0.051 i & 0.631 \\
	\end{array}
	\right)$ 
	$\pm \left(
	\begin{array}{cc}
	0.004 & 0.011 \\
	0.011 & 0.004 \\
	\end{array}
	\right)$, 	which corresponds to the fidelity of $ 0.9888 \pm 0.0008$.
	\item When Alice photon is projected to $|\phi_1^{+}\rangle$,
	Bob state becomes \\
	
	$\left(
	\begin{array}{cc}
	0.8815 & -0.3232 \\
	-0.3232 & 0.1185 \\
	\end{array}
	\right)$ \\
	
	and we obtained experimentally \\
	
	$\left(
	\begin{array}{cc}
	0.878 & -0.312-0.036 i \\
	-0.312+0.036 i & 0.122 \\
	\end{array}
	\right)$
	$\pm \left(
	\begin{array}{cc}
	0.002 & 0.01 \\
	0.01 & 0.002 \\
	\end{array}
	\right)$, 	which corresponds to the fidelity of  $0.9899\pm 0.0009$.
	
	\item When Alice photon is projected to  
	$|\phi_1^{-}\rangle$, Bob state becomes \\
	
	$\left(
	\begin{array}{cc}
	0.3239 & 0.4680 \\
	0.4680 & 0.6761 \\
	\end{array}
	\right)$ \\
	
	and we obtained experimentally \\
	
	$\left(
	\begin{array}{cc}
	0.329 & 0.465\, +0.009 i \\
	0.465\, -0.009 i & 0.671 \\
	\end{array}
	\right)$
	$\pm \left(
	\begin{array}{cc}
	0.004 & 0.011 \\
	0.011 & 0.004 \\
	\end{array}
	\right)$, 	which corresponds to the fidelity of $0.9957 \pm 0.0004$.
\end{enumerate}

\begin{center}
\textbf{Supplementary Note 8 - Error estimation}
\end{center}

We have considered error originated from the measurement side only, as the error on the preparation side will just shift the experimentally prepared state away from the desire state and therefore will be apparent from the reported state fidelity or the value of Bell inequality violation.

To estimate the error in our experiment, we have considered errors due to cross-talks--originated from the PBS extinction and absorption--wave-plate setting errors, wave-plates offset error, wave-plates retardance tolerance and error due to Poissonian statistics of the incoming photons. Cross-talk is considered here as the used PBSs were not perfect. To calculate the PBS extinction, we have carefully estimated the extinction ratio of the PBSs used on both sides (Alice and Bob) with their transmission and absorption for each polarizations.

The wave-plate setting error is considered as one has to switch the settings during collecting data for the estimation of the state fidelity. In the experiment, we used motorized stages to rotate the wave-plates to switch among the different settings. These mounts have repeatability of less than $0.02^{\circ}$. Therefore, for error estimation, we assumed that the wave-plates setting error has normal distribution with standard deviation of $0.02^{\circ}$.

A normally distributed offset error of $0.1^{\circ}$ in setting the wave-plates is also assumed as the zero of a given wave-plate could not be adjusted better than $0.1^{\circ}$.

The wave-plates retardance tolerance of $\frac{\lambda}{300}$ is also taken into account by assuming a normally distributed retardance error in each of the wave plate used. Note that, among all these errors, wave-plates retardance error is leading and it is over estimated as it is fixed with each wave-plate chosen for the experiment, moreover, we are carefully characterizing wave plates which we have not assumed here. Finally, we considered errors arising due to the photon counts following the Poissonian statistics.


\begin{thebibliography}{99}\vspace{0.5cm}
\Large{\textbf{References}}\vspace{0.5cm}
\normalsize

\bibitem{SW} Wehner, S., Winter, A.,
	\textit{Entropic uncertainty relations—a survey},
	New Journal of Physics \textbf{12}, 025009 (2010).
	
\bibitem{OW} Oppenheim, J., Wehner, S.,
	\textit{The uncertainty principle determines the nonlocality of quantum mechanics},
	Science, Vol. \textbf{330}, No. 6007, 1072 (2010).

\bibitem{Schrodinger} Schr\"{o}dinger, E.,
	\textit{Discussion of Probability Relations between Separated Systems},
	Math. Proc. of the Camb. Phil. Soc. \textbf{31}, 555 (1935).
	
\bibitem{Wiseman} Wiseman, H. M., Jones, S. J., Doherty, A. C.,
	\textit{Steering, Entanglement, Nonlocality, and the Einstein-Podolsky-Rosen Paradox},
	Phys. Rev. Lett. \textbf{98}, 140402 (2007).

\bibitem{Jevtic} Jevtic, S, Pusey, M. F., Jennings D., Rudolph, T,
	 \textit{Quantum Steering Ellipsoids},
	  	Phys. Rev. Lett. \textbf{113}, 020402 (2014).

\bibitem{PR} Popescu, S., Rohrlich, D.,
	\textit{Nonlocality as an axiom for quantum theory},
	Found. Phys. \textbf{24}, 379 (1994).

\bibitem{our} Ramanathan, R., Tuziemski, J., Horodecki, M., Horodecki, P.,
	\textit{No Quantum Realization of Extremal No-Signaling Boxes},
	Phys. Rev. Lett. \textbf{117}, 050401 (2016).

\bibitem{AQ} Navascu\'{e}s, M., Guryanova, Y., Hoban, M. J., Ac\'{i}n, A.,
	\textit{Almost quantum correlations},
	Nat. Comm. \textbf{6}, 6288 (2015).

\bibitem{Tsirelson} Cirel’son, B. S.,
	\textit{Quantum generalizations of Bell's inequality},
	Lett. Math. Phys. \textbf{4}, 83 (1980).

\bibitem{HJW} Hughston, L., Jozsa, R., W. Wootters,
	\textit{A complete classification of quantum ensembles},
	Phys. Lett. A \textbf{183}, 14 (1993).

\bibitem{Scarani} 	
	Y-Z. Zhen, K. T. Goh, Y-L. Zheng, W.-F. Cao, X. Wu, K. Chen, V. Scarani,
	\textit{Non-local games and optimal steering at the boundary of the quantum set},
	Phys. Rev. A \textbf{94}, 022116 (2016).

\bibitem{Cleve} Cleve, R., Hoyer, P., Toner, B., Watrous, J.,
	\textit{Consequences and Limits of Nonlocal Strategies},
	Proceedings of the 19th Annual IEEE Conference on Computational Complexity, 236--249, 		(2004).
	
\bibitem{Wehner06} Wehner, S.,
	\textit{Tsirelson bounds for generalized Clauser-Horne-Shimony-Holt inequalities},
	Phys. Rev. A \textbf{73}, 022110 (2006).

\bibitem{Brassard} Brassard, G., Broadbent, A.,  Tapp, A.,
	\textit{Quantum Pseudo-Telepathy},
	Foundations of Physics Vol. \textbf{35}, Issue 11, pp. 1877 (2005).

\bibitem{Kempe} Kempe, J., Kobayashi, H., Matsumoto, K., Toner, B.,  Vidick, T.,
	\textit{Entangled games are hard to approximate},
	Proceedings of the 49th Annual IEEE Symposium on Foundations of Computer Science, 			(2008)

\bibitem{NPA} Navascu\'{e}s, M., Pironio, S., Ac\'{i}n, A.,
	\textit{A convergent hierarchy of semidefinite programs characterizing the set of quantum 			correlations},
	New J. Phys. \textbf{10}, 073013 (2008).
	
\bibitem{NPA-dual} A. C. Dohery, Y-C. Liang, B. Toner and S. Wehner
	\textit{The quantum moment problem and bounds on entangled multi-prover games},
	Proc. of IEEE Conf. on Comp. Complexity '08, 199 (2008). 

\bibitem{Kempe2}
	Kempe, J., Regev, O., and Toner, B.,
	\textit{Unique games with entangled provers are easy},
	SIAM Journal on Computing, \textbf{39} 7, 3207 (2010). 

\bibitem{SPDC_cross}P. G. Kwiat, {\em et al.}
	\textit{Ultrabright source of polarization-entangled photons},
	Phys. Rev. A,{\bf 60}, R773(R) (1999).

\bibitem{State_prep}T.-C. Wei, {\em et al.}
	\textit{Synthesizing arbitrary two-photon polarization mixed states},
	Phys. Rev. A,{\bf 71}, 032329 (2005).
	
\bibitem{loophole-free1} B. Hensen et al., Loophole-free Bell inequality violation using 			electron spins, Nature \textbf{526}, 682686 (2015).
	
\bibitem{loophole-free2} M. Giustina et al., Significant-loophole-free test of Bell's theorem 			with entangled photons, Phys. Rev. Lett. \textbf{115}, 250401 (2015).

\bibitem{loophole-free3} L. K. Shalm et al., Strong Loophole-Free Test of Local Realism, 			Phys. Rev. Lett. \textbf{115}, 250402 (2015).\\

\end{thebibliography}

\begin{thebibliography}{99}\vspace{1cm}
\large{\textbf{Supplementary References:}}\vspace{0.5cm}
	 
	
\bibitem{MasanesSM} Masanes, Ll.,
	\textit{Extremal quantum correlations for N parties with two dichotomic observables per 			site}, Preprint at https://arxiv.org/abs/quant-ph/0512100 (2005).
	
\bibitem{JordanSM} Jordan, C.,
	\textit{Essai sur la geometrie a n dimensions},
	Bulletin de la S. M. F. \textbf{3}, 103 (1875).
	
\bibitem{HJWSM} Hughston, L., Jozsa, R., Wootters, W.,
	\textit{A complete classification of quantum ensembles having a given density matrix},
	Phys. Lett. A \textbf{183}, 14 (1993).
	
\bibitem{SchrodingerSM} Schr\"{o}dinger, E.,
	\textit{Discussion of Probability Relations between Separated Systems},
	Math. Proc. of the Camb. Phil. Soc. \textbf{31}, 555 (1935).
	
\bibitem{OWSM} Oppenheim, J., Wehner, S.,
	\textit{The Uncertainty Principle Determines the Nonlocality of Quantum Mechanics},
	Science, Vol. \textbf{330}, No. 6007, 1072 (2010).

\bibitem{ScaraniSM}
	Zhen, Y.-Z.,  Tong Goh, K. ,  Zheng, Y.-L., Cao, W.-F., Wu, X., Chen, K., Scarani, V.,
	\textit{Non-local games and optimal steering at the boundary of the quantum set},
	Phys. Rev. A \textbf{94}, 022116 (2016).

\bibitem{UnbSM} 
	Perez-Garcia, D., Wolf, M.M., Palazuelos, C., Villanueva, I., Junge, M., 
	\textit{Unbounded violation of tripartite Bell inequalities},
	Comm. Math. Phys. 279, 455 (2008).	
	
\bibitem{UGCSM}
	Khot, S., 
	\textit{On the power of unique 2-prover 1-round games}, 
	Proceedings of the thirty-fourth annual ACM Symposium on Theory of computing, 767,	(2002).
	
\bibitem{SMA07SM} Silman, J., Machnes, S., Aharon, N., \textit{On the relation between Bell inequalities and nonlocal games}, 	Phys. Lett. A \textbf{372}, 3796 (2008). 

\bibitem{tiltCHSHSM} Ac\'{i}n, A., Massar, S., Pironio, S.,
	\textit{Randomness versus Nonlocality and Entanglement},
	Phys. Rev. Lett. \textbf{108}, 100402 (2012).
	
\bibitem{CGS16} Coladangelo, A., Goh, K. T., Scarani, V., \textit{All Pure Bipartite Entangled States can be Self-Tested}, Nat. Comm. \textbf{8}, 15485 (2017).	

\bibitem{SPDC_crossSM} Kwiat, P. G., {\em et al.}
	\textit{Ultrabright source of polarization-entangled photons},
	{\it Phys. Rev. A},{\bf 60}, R773(R) (1999).
		
\bibitem{State_prepSM} Wei, T.-C. {\em et al.}
	\textit{Synthesizing arbitrary two-photon polarization mixed states},
	Phys. Rev. A,{\bf 71}, 032329 (2005).
	
\bibitem{tomoSM} Altepeter, J. B.,  Jeffrey, E. R., Kwiat, P. G.
	\textit{Photonic state tomography},
	Advances in Atomic, Molecular, and Optical Physics, Elsevier, {\bf 52}, 105--159, (2005)

\end{thebibliography}
\end{document}